\documentclass[11pt,draftclsnofoot,peerreview]{IEEEtran}
\pdfoutput=1

\usepackage{times}
\usepackage{graphicx}
\usepackage{amsmath}
\usepackage{amsfonts}
\usepackage{amssymb}
\usepackage{hyperref}

\newtheorem{lemma}{Lemma}[section]
\newtheorem{theorem}{Theorem}[section]
\newtheorem{corollary}{Corollary}[section]

\title{\bf{\Large Secret Sharing over Fast-Fading MIMO Wiretap
    Channels}\thanks{Tan F. Wong and John M. Shea are with the
    Wireless Information Networking Group, University of Florida,
    Gainesvilles, Florida, 32611-6130, USA. Matthieu Bloch is with the
    School of Electrical and Computer Engineering, Georgia Institute
    of Technology, Atlanta, GA, and with the GT-CNRS UMI 2958, 2-3 Rue
    Marconi, 57070, Metz, France}}

\author{Tan F. Wong, Matthieu Bloch, and John M. Shea} 

\begin{document}
\maketitle
\setcounter{page}{1}

\begin{abstract}
  Secret sharing over the fast-fading MIMO wiretap channel is
  considered. A source and a destination try to share secret
  information over a fast-fading MIMO channel in the presence of an
  eavesdropper who also makes channel observations that are different
  from but correlated to those made by the destination. An
  interactive, authenticated public channel with unlimited capacity is
  available to the source and destination for the secret sharing
  process. This situation is a special case of the ``channel model
  with wiretapper'' considered by Ahlswede and Csisz\'{a}r. An
  extension of their result to continuous channel alphabets is
  employed to evaluate the key capacity of the fast-fading MIMO
  wiretap channel. The effects of spatial dimensionality provided by
  the use of multiple antennas at the source, destination, and
  eavesdropper are then investigated.
\end{abstract}

\section{Introduction}
The wiretap channel considered in the seminal paper~\cite{Wyner1975}
is the first example that demonstrates the possibility of secure
communications at the physical layer. It is shown in~\cite{Wyner1975}
that a source can transmit a message at a positive (secrecy) rate to a
destination in such a way that an eavesdropper only gathers
information at a negligible rate, when the source-to-eavesdropper
channel\footnote{The source-to-eavesdropper and source-to-destination
  channels will hereafter be referred to as eavesdropper and
  destination channels, respectively.} is a degraded version of the
source-to-destination channel. A similar result for the Gaussian
wiretap channel is provided in~\cite{Leung1978}. The work
in~\cite{Csiszar1978} further removes the degraded wiretap channel
restriction showing that positive secrecy capacity is possible if the
destination channel is ``more capable'' (``less noisy'' for a full
extension of the rate region in~\cite{Wyner1975}) than the
eavesdropper's channel. Recently, there has been a flurry of interest
in extending these early results to more sophisticated channel models,
including fading wiretap channels, multi-input multi-output (MIMO)
wiretap channels, multiple-access wiretap channels, broadcast wiretap
channels, relay wiretap channels, \emph{etc}. We do not attempt to
provide a comprehensive summary of all recent developments, and
highlight only results that are most relevant to the present work. We
refer interested readers to the introduction and reference list
of~\cite{Liang2008} for a concise and extensive overview of recent
works.

When the destination and eavesdropper channels experience independent
fading, the strict requirement of having a more capable destination
channel for positive secrecy capacity can be loosened. This is due to
the simple observation that the destination channel may be more
capable than the eavesdropper's channel under some fading
realizations, even if the destination is not more capable than the
eavesdropper on average. Hence, if the channel state information (CSI)
of both the destination and eavesdropper channels is available at the
source, it is shown in~\cite{Liang2008,Gopala2007} that a positive
secrecy capacity can be achieved by means of appropriate power control
at the source. The key idea is to opportunistically transmit only
during those fading realizations for which the destination channel is
more capable~\cite{Bloch2008}. For block-ergodic fading, it is also
shown in~\cite{Gopala2007} (see also~\cite{Khisti2008}) that a
positive secrecy capacity can be achieved with a variable-rate
transmission scheme without any eavesdropper CSI available at the
source.

When the source, destination, and eavesdropper have multiple antennas,
the resulting channel is known as a MIMO wiretap channel
(see~\cite{Shafiee2007,Khisti2007,Oggier2007,Liu2009,Bustin2009}),
which may also have positive secrecy capacity. Since the MIMO wiretap
channel is not degraded, the characterization of its secrecy capacity
is not straightforward. For instance, the secrecy capacity of the MIMO
wiretap channel is characterized in~\cite{Khisti2007} as the saddle
point of a minimax problem, while an alternative characterization
based on a recent result for multi-antenna broadcast channels is
provided in~\cite{Liu2009}. Interestingly all characterizations point
to the fact that the capacity achieving scheme is one that transmits
only in the directions in which the destination channel is more
capable than the eavesdropper's channel. Obviously, this is only
possible when the destination and eavesdropper CSI is available at the
source. It is shown in~\cite{Khisti2007} that if the individual
channels from antennas to antennas suffer from independent Rayleigh
fading, and the respective ratios of the numbers of source and
destination antennas to that of eavesdropper antennas are larger than
certain fixed values, then the secrecy capacity is positive with
probability one when the numbers of source, destination, and
eavesdropper antennas become very large.

As discussed above, the availability of destination (and eavesdropper)
CSI at the source is an implicit requirement for positive secrecy
capacity in the fading and MIMO wiretap channels. Thus, an
authenticated feedback channel is needed to send the CSI from the
destination back to the source. In~\cite{Gopala2007,Khisti2008}, this
feedback channel is assumed to be public, and hence the destination
CSI is also available to the eavesdropper. In addition, it is assumed
that the eavesdropper knows its own CSI. With the availability of a
feedback channel, if the objective of having the source send secret
information to the destination is relaxed to distilling a secret key
shared between the source and destination, it is shown
in~\cite{Maurer1993} that a positive key rate is achievable when the
destination and eavesdropper channels are two conditionally
independent (given the source input symbols) memoryless binary
channels, even if the destination channel is not more capable than the
eavesdropper's channel. This notion of secret sharing is formalized
in~\cite{Ahlswede1993} based on the concept of \emph{common
  randomness} between the source and destination. Assuming the
availability of an interactive, authenticated public channel with
unlimited capacity between the source and destination,
\cite{Ahlswede1993} suggests two different system models, called the
``source model with wiretapper'' (SW) and the ``channel model with
wiretapper'' (CW). The CW model is the similar to the (discrete
memoryless) wiretap channel model that we have discussed before. The
SW model differs in that the random symbols observed at the source,
destination, and eavesdropper are realizations of a discrete
memoryless source with multiple components. Both SW and CW models have
been extended to the case of secret sharing among multiple terminals,
with the possibility of some terminals acting as
helpers~\cite{Csiszar2000,Csiszar2004,Csiszar2008}. Key capacities
have been obtained for the two special cases in which the
eavesdropper's channel is a degraded version of the destination
channel and in which the destination and eavesdropper channels are
conditionally independent~\cite{Ahlswede1993,Maurer1993}. Similar
results have been derived for multi-terminal secret
sharing~\cite{Csiszar2004,Csiszar2008}, with the two special cases
above subsumed by the more general condition that the terminal symbols
form a Markov chain on a tree. Authentication of the public channel
can be achieved by the use of an initial short key and then a small
portion of the subsequent shared secret message~\cite{Bennett1995}. A
detailed study of secret sharing over an unauthenticated public
channel is given in~\cite{Maurer2003,Maurer2003a,Maurer2003b}.

Other approaches to employ feedback have also been recently
considered~\cite{Lai2007,Ekrem2008CISS,Tekin2008}. In particular, it
is shown in~\cite{Lai2007} that positive secrecy capacity can be
achieved for the modulo-additive discrete memoryless wiretap channel
and the modulo-$\Lambda$ channel if the destination is allowed to send
signals back to the source over the same wiretap channel and both
terminals can operate in full-duplex manner. In fact, for the former
channel, the secrecy capacity is the same as the capacity of such a
channel in the absence of the eavesdropper.

In this paper, we consider secret sharing over a fast-fading MIMO
wiretap channel. Thus, we are interested in the CW model
of~\cite{Ahlswede1993} with memoryless conditionally independent
destination and eavesdropper channels and continuous channel
alphabets. We provide an extension of the key capacity result
in~\cite{Ahlswede1993} for this case to include continuous channel
alphabets (Theorem~\ref{thm:keycapgen}).  Using this result, we obtain
the key capacity of the fast-fading MIMO wiretap channel
(Section~\ref{se:fading}). Our result indicates that the key capacity
is always positive, no matter how large the channel gain of the
eavesdropper's channel is; in addition this holds even if the
destination and eavesdropper CSI is available at the destination and
eavesdropper, respectively. Of course, the availability of the public
channel implies that the destination CSI could be fed back to the
source. However, due to the restrictions imposed on the secret-sharing
strategies (see Section~\ref{se:secret}), only causal feedback is
allowed, and thus any destination CSI available at source is
``outdated''. This does not turn out to be a problem since, unlike the
approaches mentioned above, the source does not use the CSI to avoid
sending secret information when the destination is not more capable
than the eavesdropper's channel. As a matter of fact, the fading
process of the destination channel provides a significant part of the
common randomness from which the source and the destination distill a
secret key. This fact is readily obtained from the alternative
achievability proof given in Section~\ref{se:proof}. We note
that~\cite{KhistiISIT08,Prabhakaran2008} consider the problem key
generation from common randomness over wiretap channels and exploit a
Wyner-Ziv coding scheme to limit the amount of information conveyed
from the source to the destination via the wiretap channel. Unlike
these previous works, we only employ Wyner-Ziv coding to quantize the
destination channel outputs. Our code construction still relies on a
public channel with unlimited capacity to achieve the key capacity.

Finally, we also investigate the limiting value of the key capacity under three asymptotic scenarios. In the first
scenario, the transmission power of the source becomes asymptotically high (Corollary~\ref{thm:keycaplimP}). In the
second scenario, the destination and eavesdropper have a large number of antennas (Corollary~\ref{thm:keycapant}). In
the third scenario, the gain advantage of the eavesdropper's channel becomes asymptotically large
(Corollary~\ref{thm:keycaplim}). These three scenarios reveal two different effects of spatial dimensionality upon key
capacity. In the first scenario, we show that the key capacity levels off as the power increases if the eavesdropper has
no fewer antennas than the source. On the other hand, when the source has more antennas, the key capacity can increase
without bound with the source power. In the second scenario, we show that the spatial dimensionality advantage that the
eavesdropper has over the destination has exactly the same effect as the channel gain advantage of the eavesdropper.  In
the third scenario, we show that the limiting key capacity is positive only if the eavesdropper has fewer antennas than
the source. The results in these scenarios confirm that spatial dimensionality can be used to combat the eavesdropper's
gain advantage, which was already observed for the MIMO wiretap channel. Perhaps more surprisingly, this is achieved
with neither the source nor destination needing any eavesdropper CSI.

\section{Secret Sharing and Key Capacity}
\label{se:secret}

We consider the CW model of~\cite{Ahlswede1993}, and we recall its characteristics for completeness. We consider three
terminals, namely a source, a destination, and an eavesdropper. The source sends symbols from an alphabet
$\mathcal{X}$. The destination and eavesdropper observe symbols belonging to alphabets $\mathcal{Y}$ and $\mathcal{Z}$,
respectively. Unlike in~\cite{Ahlswede1993}, $\mathcal{X}$, $\mathcal{Y}$, and $\mathcal{Z}$ need not be discrete. In
fact, in Section~\ref{se:fading} we will assume they are multi-dimensional vector spaces over the complex field. The
channel from the source to the destination and eavesdropper is assumed memoryless. A generic symbol sent by the source
is denoted by $X$ and the corresponding symbols observed by the destination and eavesdropper are denoted by $Y$ and $Z$,
respectively. For notational convenience (and without loss of generality), we assume that $(X,Y,Z)$ are jointly
continuous, and the channel is specified by the conditional probability density function (pdf) $p_{Y,Z|X}(y,z|x)$. In
addition, we restrict ourselves to cases in which $Y$ and $Z$ are conditionally independent given $X$, i.e.,
$p_{Y,Z|X}(y,z|x) = p_{Y|X}(y|x) p_{Z|X}(z|x)$, which is a reasonable model for symbols broadcasted in a wireless
medium. Hereafter, we drop the subscripts in pdfs whenever the concerned symbols are well specified by the arguments of
the pdfs. We assume that an interactive, authenticated public channel with unlimited capacity is also available for
communicatin between the source and destination. Here, \emph{interactive} means that the channel is two-way and can be
used multiple times, \emph{unlimited capacity} means that it is noiseless and has infinite capacity, and \emph{public}
and \emph{authenticated} mean that the eavesdropper can perfectly observe all communications over this channel but
cannot tamper with the messages transmitted.

We consider the class of permissible secret-sharing strategies suggested in~\cite{Ahlswede1993}. Consider $k$ time
instants labeled by $1,2,\ldots,k$, respectively. The $(X,Y,Z)$ channel is used $n$ times during these $k$ time instants
at $i_1 < i_2 < \cdots < i_n$. Set $i_{n+1}=k$. The public channel is used for the other ($k-n$) time instants. Before
the secret-sharing process starts, the source and destination generate, respectively, independent random variable $M_X$
and $M_Y$.  To simplify the notation, let $a^i$ represent a sequence of messages/symbols $a_1, a_2, \ldots, a_i$.  Then
a permissible strategy proceeds as follows:
\begin{itemize}
\item At time instant $0<i<i_1$, the source sends message
  $\Phi_i=\Phi_i(M_X,\Psi^{i-1})$ to the destination, and the
  destination sends message $\Psi_i=\Psi_i(M_Y,\Phi^{i-1})$ to the
  source. Both transmissions are carried over the public channel.
\item At time instant $i=i_j$ for $j=1,2,\ldots,n$, the source sends
  the symbol $X_j=X_j(M_X,\Psi^{i_j-1})$ to the $(X,Y,Z)$ channel. The
  destination and eavesdropper observe the corresponding symbols $Y_j$
  and $Z_j$. There is no message exchange via the public channel,
  i.e., $\Phi_i$ and $\Psi_i$ are both null.
\item At time instant $i_j < i < i_{j+1}$ for $j=1,2,\ldots,n$, the
  source sends message $\Phi_i=\Phi_i(M_X,\Psi^{i-1})$ to the
  destination, and the destination sends message
  $\Psi_i=\Psi_i(M_Y,Y^{j},\Phi^{i-1})$ to the source. Both
  transmissions are carried over the public channel.
\end{itemize}
At the end of the $k$ time instants, the source generates its secret
key $K=K(M_X,\Psi^k)$, and the destination generates its secret key
$L=L(M_Y,Y^n,\Phi^k)$, where $K$ and $L$ takes values from the same
finite set $\mathcal{K}$.

According to~\cite{Ahlswede1993}, $R$ is an \emph{achievable key rate}
through the channel $(X,Y,Z)$ if for every $\varepsilon>0$, there
exists a permissible secret-sharing strategy of the form described
above such that
\begin{enumerate}
\item $\Pr\{K\neq L\} < \varepsilon$,
\item $\frac{1}{n} I(K;Z^n, \Phi^k, \Psi^k) < \varepsilon$,
\item $\frac{1}{n} H(K) > R - \varepsilon$, and
\item $\frac{1}{n} \log |\mathcal{K}| < \frac{1}{n} H(K) + \varepsilon$,
\end{enumerate}
for sufficiently large $n$. The \emph{key capacity} of the channel $(X,Y,Z)$ is the largest achievable key rate through
the channel. We are interested in finding the key capacity.  For the case of continuous channel alphabets considered
here, we also add the following power constraint to the symbol sequence $X^n$ sent out by the source:
\begin{equation}
\frac{1}{n} \sum_{j=1}^{n} |X_j|^2 \leq P
\label{e:powerconstr}
\end{equation}
with probability one (w.p.1) for sufficiently large $n$.

\begin{theorem}\label{thm:keycapgen}
  The key capacity of a CW model $(X,Y,Z)$ with conditional pdf $p(y,z|x)=p(y|x)p(z|x)$ is given by $\max_{X: E[|X|^2]
    \leq P} [I(X;Y) - I(Y;Z)]$.
\end{theorem}
\begin{proof}
  The case with discrete channel alphabets is established in
  \cite[Corollary 2 of Theorem 2]{Ahlswede1993}, whose achievability
  proof (also the ones in~\cite{Csiszar2004,Csiszar2008}) does not
  readily extend to continuous channel alphabets.  Nevertheless the
  same single backward message strategy suggested in
  ~\cite{Ahlswede1993} is still applicable for continuous
  alphabets. That strategy uses $k=n+1$ time instants with $i_j=j$ for
  $j=1,2,\ldots,n$. That is the source first sends $n$ symbols through
  the $(X,Y,Z)$ channel; after receiving these $n$ symbols, the
  destination feeds back a single message at the last time instant to
  the source over the public channel.  A carefully structured
  Wyner-Ziv code can be employed to support this secret-sharing
  strategy. The detailed arguments are provided in the alternative
  achievability proof in Section~\ref{se:proof}.

  Here we outline an achievability argument based on the consideration of a conceptual wiretap channel from the
  destination back to the source and eavesdropper suggested in \cite[Theorem~3]{Maurer1993}.  First, assume the source
  sends a sequence of i.i.d. symbols $X^n$, each distributed according to $p(x)$, over the wiretap channel. Suppose that
  $E[|X|^2] \leq P$. Because of the law of large numbers, we can assume that $X^n$ satisfies the power constraint
  (\ref{e:powerconstr}) without loss of generality. Let $Y^n$ and $Z^n$ be the observations of the the destinations and
  eavesdropper, respectively. To transmit a sequence $U^n$ of symbols independent of $(X^n,Y^n,Z^n)$, the destination
  sends $U^n+Y^n$ back to the source via the public channel. This creates a conceptual memoryless wiretap channel from
  the destination with input symbol $U$ to the source in the presence of the eavesdropper, where the source observes
  $(U+Y,X)$ while the eavesdropper observes $(U+Y,Z)$.

  Employing the continuous alphabet extension of the well known result in~\cite{Csiszar1978}, the secrecy capacity of
  the conceptual wiretap channel (and hence the key capacity of the original channel) is lower bounded by
  \[
  \max_U [ I(U;U+Y,X) - I(U;U+Y,Z)].
  \]
  Note that the input symbol $U$ has no power constraint since the
  public channel has infinite capacity.  But
\begin{eqnarray}
\lefteqn{I(U;U+Y,X) - I(U;U+Y,Z)} \nonumber \\
&=& 
I(U;X) + I(U;U+Y|X) - [I(U;Z) + I(U;U+Y|Z)] \nonumber \\
&=&
h(U) - h(U|X) + h(U+Y|X) - h(U+Y|U,X) - h(U) + h(U|Z) - h(U+Y|Z) +
h(U+Y|U,Z)  \nonumber \\
&=&
h(Y|Z) - h(Y|X) + [h(U+Y|X) - h(U|X)] - [h(U+Y|Z) - h(U|Z)] \nonumber \\
&\geq&
h(Y|Z) - h(Y|X) - [h(U+Y|X) - h(U|X)]  \nonumber \\
&\geq&
h(Y|Z) - h(Y|X) - [h(U+Y) - h(U)] 
\label{e:keycaplb}
\end{eqnarray}
where the equality on the fourth line results from $h(U+Y|U,X) =
h(Y|U,X) = h(Y|X)$ due to the independence of $U$ and $Y$,
the inequality on the fifth line follows from 
the fact 
\[
h(U+Y|Z)-h(U|Z) \geq h(U+Y|Z,Y) - h(U|Z) = h(U|Z,Y) - h(U|Z) = 0,
\]
which is again due to independence between $(Y,Z)$ and $U$, and the
inequality on the last line follows from $h(U+Y|X)-h(U|X) = h(U+Y|X) -
h(U) \leq h(U+Y) - h(U)$.

Without loss of generality and for notational simplicity, assume that $Y$ and $U$ are both one-dimensional real random
variables. Now, choose $U$ to be Gaussian distributed with mean $0$ and variance $\sigma_U^2$. Then
\begin{eqnarray}
h(U+Y) - h(U) &\leq& 
 \frac{1}{2}\log \left(2\pi e \mathrm{var}(U+Y)\right) - \frac{1}{2}
 \log (2\pi e \sigma_U^2) \nonumber \\
 &=&
 \frac{1}{2} \log \left( \frac{\sigma^2_U +
     \mathrm{var}(Y)}{\sigma^2_U} \right)
 \label{e:hU+Y-hU}
\end{eqnarray}
where the first inequality follows from~\cite[Theorem 8.6.5]{Cover2006} and the last equality is due to the independence
between $Y$ and $U$. Combining (\ref{e:keycaplb}) and (\ref{e:hU+Y-hU}), for every $\varepsilon >0$, we can choose
$\sigma^2_U$ large enough such that
\[
I(U;U+Y,X) - I(U;U+Y,Z) \geq h(Y|Z)-h(Y|X) - \varepsilon =
I(X;Y)-I(Y;Z) - \varepsilon.
\]
Since $\varepsilon$ is arbitrary, the key capacity is lower bounded by $\max_{E[|X|^2] \leq P} [I(X;Y)-I(Y;Z)]$.

The converse proof in~\cite{Ahlswede1993} is directly applicable to continuous channel alphabets, provided the average
power constraint (\ref{e:powerconstr}) can be incorporated into the arguments in
\cite[pp. 1129--1130]{Ahlswede1993}. This latter requirement is simplified by the additive and symmetric nature of the
average power constraint \cite[Section 3.6]{Han2003}. To avoid too much repetition, we outline below only the steps of
the proof that are not directly available in \cite[pp. 1129--1130]{Ahlswede1993}.

For every permissible strategy with achievable key rate $R$, we have
\begin{eqnarray}
\frac{1}{n} I(K;L) 
&=& \frac{1}{n}H(K) - \frac{1}{n}H(K|L) \nonumber \\
&\geq& 
\frac{1}{n}H(K) - \frac{1}{n} \left[1 + \Pr\{K\neq L\} \cdot 
  \log |\mathcal{K}|\right]
  \nonumber \\
&>&
\frac{1}{n}H(K) - \frac{1}{n} - \varepsilon \left[ \frac{1}{n} H(K) + \varepsilon \right] 
\nonumber \\
&>& (1-\varepsilon) (R-\varepsilon) - \frac{1}{n} - \varepsilon^2
\label{e:IKL}
\end{eqnarray}
where the second line follows from Fano's inequality, the third line results from conditions 1) and 4) in the definition
of achievable key rate, and the last line is due to condition 3). Thus it suffices to upper bound $I(K;L)$.  From
condition 2) in the definition of achievable key rate and the chain rule, we have
\begin{eqnarray}
\frac{1}{n} I(K;L) &<&
 \frac{1}{n} I(K;L|Z^n,\Phi^k,\Psi^k) + \varepsilon \nonumber \\
&\leq&
  \frac{1}{n} I(M_X ; M_Y, Y^n |Z^n,\Phi^k,\Psi^k) + \varepsilon 
\label{e:ubIKL}
\end{eqnarray}
where the second inequality is due to the fact that $K=K(M_X,\Psi^k)$ and $L=L(M_Y,Y^n,\Phi^k)$. By repeated uses of the
chain rule, the construction of permissible strategies, and the memoryless nature of the $(X,Y,Z)$ channel, it is shown
in \cite[pp. 1129--1130]{Ahlswede1993} that
\begin{equation}
  \frac{1}{n} I(M_X ; M_Y, Y^n |Z^n,\Phi^k,\Psi^k) 
  \leq
  \frac{1}{n} \sum_{j=1}^{n} I(X_j;Y_j|Z_j).
\label{e:IXY|Z}
\end{equation}

Now let $Q$ be a uniform random variable that takes value from $\{1,2,\ldots,n\}$, and is independent of all other
random quantities. Define $(\tilde X, \tilde Y, \tilde Z) = (X_j, Y_j, Z_j)$ if $Q=j$. Then it is obvious that
$p_{\tilde Y, \tilde Z| \tilde X}(\tilde y, \tilde z|\tilde x)=p_{Y,Z|X}(\tilde y, \tilde z|\tilde x)$, and
(\ref{e:IXY|Z}) can be rewritten as
\begin{equation}
  \frac{1}{n} I(M_X ; M_Y, Y^n |Z^n,\Phi^k,\Psi^k) 
  \leq I(\tilde X; \tilde Y | \tilde Z, Q)
  \leq I(\tilde X; \tilde Y | \tilde Z)
\label{e:ItXY|Z}
\end{equation}
where the second inequality is due to the fact that $Q \rightarrow \tilde X \rightarrow (\tilde Y,\tilde Z)$ forms a
Markov chain.  On the other hand, the power constraint (\ref{e:powerconstr}) implies that
\begin{equation}
  E[|\tilde X|^2] =  \frac{1}{n} \sum_{j=1}^{n} E[|X_j|^2] \leq P.
\label{e:avpwconstr}
\end{equation}

Combining (\ref{e:IKL}), (\ref{e:ubIKL}), and (\ref{e:ItXY|Z}), we obtain
\begin{equation}
  R < \frac{1}{1-\varepsilon} \left[
    I(\tilde X;\tilde Y|\tilde Z) + 2\varepsilon + \frac{1}{n}
  \right].
 \label{e:ubR}
\end{equation}
Since $\varepsilon$ can be arbitrarily small when $n$ is sufficiently
large, (\ref{e:ubR}), together with (\ref{e:avpwconstr}), gives
\begin{eqnarray*}
  R &\leq& I(\tilde X;\tilde Y|\tilde Z) \\ 
  &\leq& \max_{X:E[|X|^2]\leq P} I(X;Y|Z) \\
  &=& \max_{X:E[|X|^2]\leq P} [I(X;Y)-I(Y;Z)]
\end{eqnarray*}
where the last line is due to the fact that $p(y,z|x)=p(y|x)p(z|x)$.
\end{proof}

\section{Key Capacity of Fast Fading MIMO Wiretap Channel}
\label{se:fading}

Consider that the source, destination, and eavesdropper have $m_S$,
$m_D$, and $m_W$ antennas, respectively. The antennas in each node are
separated by at least a few wavelengths, and hence the fading
processes of the channels across the transmit and receive antennas are
independent.  Using the complex baseband representation of the
bandpass channel model:
\begin{eqnarray}
Y_D &=& H_D X + N_D \nonumber \\
Y_W &=& \alpha H_W X + N_W
\label{e:model}
\end{eqnarray}
where
\begin{itemize}
\item $X$ is the $m_S\times 1$ complex-valued transmit symbol vector
  by the source,
\item $Y_D$ is the $m_D\times 1$ complex-valued receive symbol vector
  at the destination,
\item $Y_W$ is the $m_W\times 1$ complex-valued receive symbol vector
  at the eavesdropper,
\item $N_D$ is the $m_D\times 1$ noise vector with independent
  identically distributed (i.i.d.) zero-mean, circular-symmetric
  complex Gaussian-distributed elements of variance $\sigma_D^2$
  (i.e., the real and imaginary parts of each elements are independent
  zero-mean Gaussian random variables with the same variance),
\item $N_W$ is the $m_W\times 1$ noise vector with i.i.d. zero-mean,
  circular-symmetric complex Gaussian-distributed elements of variance
  $\sigma_W^2$,
\item $H_D$ is the $m_D\times m_S$ channel matrix from the source to
  destination with i.i.d. zero-mean, circular-symmetric complex
  Gaussian-distributed elements of unit variance,
\item $H_W$ is the $m_W\times m_S$ channel matrix from the source to
  eavesdropper with i.i.d. zero-mean, circular-symmetric complex
  Gaussian-distributed elements of unit variance
\item $\alpha>0$ models the gain advantage of the eavesdropper over
  the destination.
\end{itemize}
Note that $H_D$, $H_W$, $N_D$, and $N_W$ are independent.  The wireless channel modeled by (\ref{e:model}) is used $n$
times as the $(X,Y,Z)$ channel described in Section~\ref{se:secret} with $Y=[Y_D \ H_D]$ and $Z=[Y_W \ H_W]$. We assume
that the $n$ uses of the wireless channel in (\ref{e:model}) are i.i.d. so that the memoryless requirement of the
$(X,Y,Z)$ channel is satisfied.  Since $H_D$ and $H_W$ are included in the respective channel symbols observable by the
destination and eavesdropper (i.e., $Y$ and $Z$ respectively), this model also implicitly assumes that the destination
and eavesdropper have perfect CSI of their respective channels from the source. In practice, we can separate adjacent
uses of the wireless channel by more than the coherence time of the channel to approximately ensure the i.i.d. channel
use assumption. Training (known) symbols can be sent right before or after (within the channel coherence period) by the
source so that the destination can acquire the required CSI. The eavesdropper may also use these training symbols to
acquire the CSI of its own channel. If the CSI required at the destination is obtained in the way just described, then a
unit of channel use includes the symbol $X$ together with the associated training symbols. However, as
in~\cite{Telatar1999}, we do not count the power required to send the training symbols (cf. Eq.~(\ref{e:powerconstr})).
Moreover we note that the source (and also the eavesdropper) may get some information about the outdated CSI of the
destination channel, because information about the destination channel CSI, up to the previous use, may be fed back to
the source from the destination via the public channel. More specifically, at time instant $i_j$, the source symbol
$X_j$ is a function of the feedback message $\Psi^{i_j-1}$, which is in turn some function of the realizations of $H_D$
at time $i_1, i_2, \ldots, i_{j-1}$. We also note that neither the source nor destination has any eavesdropper CSI.
Referring back to (\ref{e:model}), these two facts imply that $X$ is independent of $H_D$, $H_W$, $N_D$, and $N_W$,
i.e., the current source symbol $X$ is independent of the current channel state.

Since the fading MIMO wiretap channel model in (\ref{e:model}) is a
special case of the CW model considered in Section~\ref{se:secret},
the key capacity $C_K$ is given by Theorem~\ref{thm:keycapgen} as:
\begin{equation}
C_K = \max_{X: E[|X|^2] \leq P} [I(X;Y_D, H_D) - I(Y_D, H_D;Y_W, H_W)].
\label{e:CK1}
\end{equation}
Note that
\begin{eqnarray}
I(X; Y_D, H_D) - I(Y_D, H_D ; Y_W, H_W) 
&=& I(X; Y_D | H_D) - I (Y_D; Y_W | H_D, H_W) \nonumber \\
&=& h(Y_D|Y_W, H_D, H_W) - h(Y_D|X,H_D) \nonumber \\
&=& h(Y_D|Y_W, H_D, H_W) - m_D \log(\pi e \sigma_D^2).
\label{e:keyrate}
\end{eqnarray}
Substituting this back into (\ref{e:CK1}), we get
\begin{equation}
C_K =
 \max_{X:E[|X|^2] \leq P} h(Y_D|Y_W, H_D, H_W) - m_D \log(\pi e \sigma_D^2).
\label{e:CK2}
\end{equation}
As a result, the key capacity of the fast-fading wiretap channel
described by (\ref{e:model}) can be obtained by maximizing the
conditional entropy $h(Y_D|Y_W, H_D, H_W)$. This maximization problem
is solved below:
\begin{theorem} \label{thm:keycap}
 \[
 C_K = E \left[ \log \frac{\det \left(I_{m_S} + \frac{\alpha^2 P}{m_S
         \sigma_W^2} H_W^{\dagger} H_W + \frac{P}{m_S \sigma_D^2}
       H_D^{\dagger} H_D \right)}{\det \left(I_{m_S} + \frac{\alpha^2
         P}{m_S \sigma_W^2} H_W^{\dagger} H_W \right)} \right].
 \]
 where $\dagger$ denotes conjugate transpose.
\end{theorem}
\begin{proof}
  To determine the key capacity, we need the following upper bound on
  the conditional entropy $h(U|V)$
\begin{lemma} \label{thm:condentropy} Let $U$ and $V$ be two jointly
  distributed complex random vectors of dimensions $m_U$ and $m_V$,
  respectively. Let $K_U$, $K_V$, and $K_{UV}$ be the covariance of
  $U$, covariance of $V$, and cross-covariance of $U$ and $V$,
  respectively. If $K_V$ is invertible, then
\[
h(U|V) \leq  
 \log\det(K_U - K_{UV} K_V^{-1} K_{VU}) + m_U \log (\pi e).
\]
The upper bound is achieved when $[U^{T}\, V^{T}]^{T}$ is a
circular-symmetric complex Gaussian random vector.
\end{lemma}
\begin{proof}
  We can assume that both $U$ and $V$ have zero means without loss of
  generality. Also assume that the existence of all unconditional and
  conditional covariances stated below.  For each $v$,
\begin{equation}
h(U|V=v) \leq \log \left( (\pi e)^{m_U} \det(K_{U|v}) \right)
\label{e:hUV=v}
\end{equation}
where $K_{U|v}$ is the covariance of $U$ with respect to the
conditional density $p_{U|V}(u|v)$ \cite[Lemma 2]{Telatar1999}.  This
implies
\begin{eqnarray}
h(U|V) &\leq & 
 E_V \left[ \log \left( (\pi e)^{m_U} \det(K_{U|V}) \right)\right]
 \nonumber \\
 & \leq &
 \log\det(E_V [K_{U|V}])  + m_U \log (\pi e) 
 \nonumber \\
 & \leq &
 \log\det(K_U - K_{UV} K_V^{-1} K_{VU}) + m_U \log (\pi e).
\label{e:hUV}
\end{eqnarray}
The second inequality above is due to the concavity of the function
$\log\det$ over the set of positive definite symmetric matrices
\cite[7.6.7]{Horn1985} and the Jensen's inequality.  To get the third
inequality, observe that $E_V [K_{U|V}]$ can be interpreted as the
covariance of the estimation error of estimating $U$ by the
conditional mean estimator $E[U|V]$. On the other hand, $K_U - K_{UV}
K_V^{-1} K_{VU}$ is the covariance of the estimation error of using
the linear minimum mean squared error estimator $K_{UV}K_V^{-1} V$
instead. The inequality results from the fact that $K_U - K_{UV}
K_V^{-1} K_{VU} \geq E_V [K_{U|V}]$ (i.e., $[K_U - K_{UV} K_V^{-1}
K_{VU}] - E_V [K_{U|V}]$ is positive semidefinite)~\cite{Scharf1990}
and the inequality of $\det(A) \geq \det(B)$ if $A$ and $B$ are
positive definite, and $A \geq B$ \cite[7.7.4]{Horn1985}.

Suppose that $[U^{T}\, V^{T}]^{T}$ is a circular-symmetric complex
Gaussian random vector. For each $v$, the conditional covariance of
$U$, conditioned on $V=v$, is the same as the (unconditional)
covariance of $U - K_{UV}K_V^{-1} V$. Since $U - K_{UV}K_V^{-1} V$ is
a circular-symmetric complex Gaussian random vector \cite[Lemma
3]{Telatar1999}, so is $U$ conditioned on $V=v$. Hence by \cite[Lemma
2]{Telatar1999}, the upper bound in (\ref{e:hUV=v}) is achieved with
$K_{U|v} = K_U - K_{UV} K_V^{-1} K_{VU}$, which also gives the upper
bound in (\ref{e:hUV}).
\end{proof}

To prove the theorem, we first obtain an upper bound on $C_K$ and then
show that the upper bound is achievable. Using
Lemma~\ref{thm:condentropy}, we have
\begin{equation}
h(Y_D|Y_W, H_D, H_W) - m_D \log(\pi e\sigma_D^2) \leq E \left[
  \log\det \left( K_{Y_D} - K_{Y_D Y_W} K_{Y_W}^{-1} K_{Y_W
      Y_D}\right) \right] - m_D \log\sigma_D^2
\label{e:ubh}
\end{equation}
where $K_{Y_D}$ and $K_{Y_W}$ are respectively the conditional
covariances of $Y_D$ and $Y_W$, given $H_D$ and $H_W$, and $K_{Y_D
  Y_W}$ and $K_{Y_W Y_D}$ are the corresponding conditional
cross-covariances.  Substituting (\ref{e:ubh}) into (\ref{e:CK2}), an
upper bound on $C_K$ is
\begin{equation}
\max_{X:E[|X|^2] \leq P} E \left[
    \log\det \left( K_{Y_D} - K_{Y_D Y_W} K_{Y_W}^{-1} K_{Y_W
        Y_D}\right)\right] - m_D \log\sigma_D^2.
\label{e:maxprob1}
\end{equation}
Thus we need to solve the maximization problem (\ref{e:maxprob1}). To
do so, let $\lambda_1, \lambda_2, \ldots, \lambda_{m_S}$ be the
(nonnegative) eigenvalues of $K_X$. Since both the distributions of
$H_D$ and $H_W$ are invariant to any unitary transformation
\cite[Lemma~5]{Telatar1999}, we can without any ambiguity define
\begin{eqnarray}
\lefteqn{f(\lambda_1, \lambda_2, \ldots, \lambda_{m_S})} \nonumber \\
&=&
  E \left[ \log\det \left( I_{m_D} + \frac{1}{\sigma_D^2} H_D K_X^{1/2}
      \left( I_{m_S} + \frac{\alpha^2}{\sigma_W^2}
        K_X^{1/2}H_W^{\dagger} H_W K_X^{1/2} \right)^{-1} K_X^{1/2}
      H_D^{\dagger} \right) \right].
\label{e:f}
\end{eqnarray}
That is, we can assume $K_X=\mathrm{diag}(\lambda_1, \lambda_2,
\ldots, \lambda_{m_S})$ with no loss of generality.  Then we have the
following lemma, which suggests that the objective function in
(\ref{e:maxprob1}) is a concave function depending only on the
eigenvalues of the covariance of $X$:
\begin{lemma}
\label{thm:Elogdet}
Suppose that $X$ has an arbitrary covariance $K_X$, whose
(nonnegative) eigenvalues are $\lambda_1, \lambda_2, \ldots,
\lambda_{m_S}$. Then 
\begin{equation}
E \left[ \log\det \left( K_{Y_D} - K_{Y_D Y_W} K_{Y_W}^{-1} K_{Y_W
  Y_D}\right) \right] - m_D \log\sigma_D^2 
= f(\lambda_1, \lambda_2, \ldots, \lambda_{m_S})
\label{e:Elogdet}
\end{equation}
is concave in $\Lambda=\{ \lambda_i \geq 0 \mbox{~for~} i=1,2,\ldots,
m_S\}$.
\end{lemma}
\begin{proof} 
  First write $A_D = H_D K_X^{1/2}$ and $A_W = \alpha H_W
  K_X^{1/2}$. It is easy to see from (\ref{e:model}) that $K_{Y_D} =
  A_D A_D^{\dagger} + \sigma^2_D I_{m_D}$, $K_{Y_W} = A_W
  A_W^{\dagger} + \sigma^2_W I_{m_W}$, and $K_{Y_D Y_W} = A_D
  A_W^{\dagger}$. Then
\begin{eqnarray}
\lefteqn{K_{Y_D} - K_{Y_D Y_W} K_{Y_W}^{-1} K_{Y_W Y_D}} \nonumber \\
&=& 
\sigma_D^2 \left\{ I_{m_D} + \frac{1}{\sigma_D^2} A_D \left[ I_{m_S}
    -  A_W^{\dagger} \left( A_WA_W^{\dagger} + \sigma_W^2 I_{m_W} \right)^{-1}
  A_W  \right] A_D^{\dagger} \right\} \nonumber \\
&=& 
\sigma_D^2 \left\{ I_{m_D} + \frac{1}{\sigma_D^2}
A_D \left[ I_{m_S} + \frac{1}{\sigma_W^2} A_W^{\dagger} A_W \right]^{-1}
A_D^{\dagger} \right\}
\label{e:KYD-}
\end{eqnarray}
where the last equality is due to the matrix inversion
formula. Substituting this result into the left hand side of
(\ref{e:Elogdet}), we obtain the right hand side of (\ref{e:f}), and
hence (\ref{e:Elogdet}).

To show concavity of $f$, it suffices to consider only diagonal
$K_X=\mbox{diag}(\lambda_1, \lambda_2, \ldots, \lambda_{m_S})$ in
$\Lambda$. Note that the mapping $H: K_X \rightarrow \left[
  \begin{array}{cc}
    K_{Y_D} & K_{Y_D Y_W} \\
    K_{Y_W Y_D} & K_{Y_W}
  \end{array}
\right]$ is linear in $\Lambda$. Also the mapping $F: \left[
  \begin{array}{cc}
    K_{Y_D} & K_{Y_D Y_W} \\
    K_{Y_W Y_D} & K_{Y_W}
  \end{array} \right]
\rightarrow  K_{Y_D} - K_{Y_D Y_W} K_{Y_W}^{-1} K_{Y_W Y_D}$ is matrix-concave
in $H(\Lambda)$ \cite[Ex. 3.58]{boyd2004co}. Thus the composition
theorem~\cite{boyd2004co} gives that the mapping $G: K_X \rightarrow
K_{Y_D} - K_{Y_D Y_W} K_{Y_W}^{-1} K_{Y_W Y_D}$ is matrix-concave in
$\Lambda$, since $G=F \circ H$. Another use of the composite theorem
together with the concavity of the function $\log\det$ as mentioned in
the proof of Lemma~\ref{thm:condentropy} shows that $\log\det G$ is
concave in $\Lambda$. Thus (\ref{e:Elogdet}) implies that $f$ is
also concave in $\Lambda$.
\end{proof}
Hence it suffices to consider only those $X$ with zero mean in
(\ref{e:maxprob1}).

Now define the constraint set $\Lambda_P=\{\lambda_i \geq 0 \mathrm{\
  for\ } i=1,2,\ldots,m_S \mathrm{\ and\ } \sum_{i=1}^{m_S} \lambda_i
\leq P\}$.  Lemma~\ref{thm:Elogdet} implies that we can find the upper
bound on $C_K$ by calculating $\max_{\Lambda_P} f(\lambda_1,
\lambda_2, \ldots, \lambda_{m_S})$, whose value is given by the next
lemma:
\begin{lemma} \label{thm:maxf} $\displaystyle \max_{\Lambda_P}
  f(\lambda_1, \lambda_2, \ldots, \lambda_{m_S}) = f
  \left(\frac{P}{m_S},\frac{P}{m_S}, \ldots, \frac{P}{m_S}\right)$.
\end{lemma}
\begin{proof} 
  Since the elements of both $H_D$ and $H_W$ are i.i.d., $f$ is
  invariant to any permutation of its arguments. This means that $f$
  is a symmetric function. By Lemma~\ref{thm:Elogdet}, $f$ is also
  concave in $\Lambda_P$. Thus it is Schur-concave
  ~\cite{Marshall1979}. Hence a Schur-minimal element (an element
  majorized by any another element) in $\Lambda_P$ maximizes $f$. It
  is easy to check that $\left(\frac{P}{m_S},\frac{P}{m_S}, \ldots,
    \frac{P}{m_S}\right)$ is Schur-minimal in $\Lambda_P$. Hence
  $\max_{\Lambda_P} f(\lambda_1, \lambda_2, \ldots, \lambda_{m_S}) = f
  \left(\frac{P}{m_S},\frac{P}{m_S}, \ldots, \frac{P}{m_S}\right)$.
\end{proof}

Combining the results in (\ref{e:maxprob1}), (\ref{e:f}),
Lemmas~\ref{thm:Elogdet} and \ref{thm:maxf}, we obtain the upper bound
on the key capacity as
\begin{eqnarray}
  C_K & \leq &
  E \left[ \log\det \left( I_{m_D} + \frac{P}{m_S\sigma_D^2}
      H_D \left(I_{m_S} + \frac{\alpha^2P}{m_S\sigma_W^2} H_W^{\dagger} H_W \right)^{-1} H_D^{\dagger} \right) \right] \nonumber \\
  &=& 
  E \left[ \log \frac{\det \left(I_{m_S} + \frac{\alpha^2 P}{m_S \sigma_W^2} H_W^{\dagger} H_W + \frac{P}{m_S \sigma_D^2} H_D^{\dagger} H_D \right)}{\det \left(I_{m_S} + \frac{\alpha^2 P}{m_S \sigma_W^2} H_W^{\dagger} H_W \right)} \right]
\label{e:CKub}
\end{eqnarray}
where the identity $\det(I+UV^{-1}U^{\dagger}) =
\frac{\det(V+U^{\dagger}U)}{\det(V)}$ for invertible $V$
\cite[Theorem 18.1.1]{Harville1997} has been used. 

On the other hand, consider choosing $X$ to have i.i.d. zero-mean,
circular-symmetric complex Gaussian-distributed elements of variance
$\frac{P}{m_S}$. Then conditioned on $H_D$ and $H_W$, $[Y_D^T \,
Y_W^T]^T$ are a circular-symmetric complex Gaussian random vector, by
applying \cite[Lemmas 3 and 4]{Telatar1999} to the linear model of
(\ref{e:model}). Hence Lemma~\ref{thm:condentropy} gives
\[
h(Y_D|Y_W, H_D, H_W) = E \left[ \log\det \left( K_{Y_D} - K_{Y_D Y_W}
    K_{Y_W}^{-1} K_{Y_W Y_D}\right) \right] + m_D \log(\pi e)
\]
where $K_{Y_D} = \frac{P}{m_S} H_D H_D^{\dagger} + \sigma^2_D
I_{m_D}$, $K_{Y_W} = \frac{\alpha^2 P}{m_S} H_W H_W^{\dagger} +
\sigma^2_W I_{m_W}$, and $K_{Y_D Y_W} = \frac{\alpha P}{m_S} H_D
H_W^{\dagger}$. Substituting this back into (\ref{e:keyrate}) and
using the matrix inversion formula to simplify the resulting
expression, we obtain the same expression on the first line of
(\ref{e:CKub}) for $I(X; Y_D, H_D) - I(Y_D, H_D ; Y_W, H_W)$. Thus the
upper bound in (\ref{e:CKub}) is achievable with this choice of $X$;
hence it is in fact the key capacity.
\end{proof}

\begin{figure}
  \begin{center}
   \includegraphics[width=0.80\textwidth]{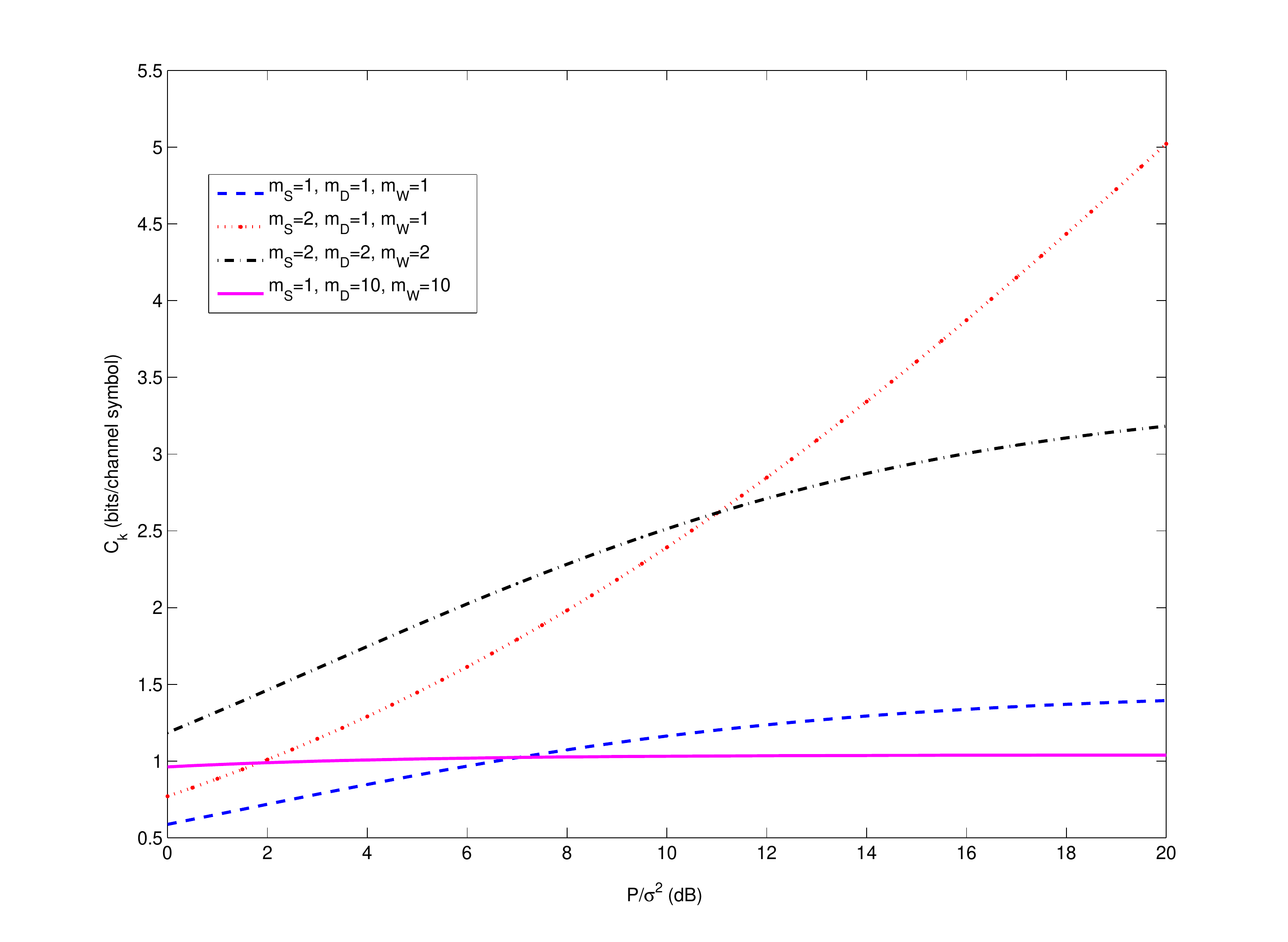}
   \caption{Key capacities of fast-fading MIMO wiretap channels with
     different numbers of source, destination, eavesdropper
     antennas. The eavesdropper's channel gain $\alpha^2=0$dB, and
     $\sigma_D^2=\sigma_W^2=\sigma^2$.}
   \vspace{-15pt}
   \label{fig1}
  \end{center}
\end{figure}
In Fig.~\ref{fig1}, the key capacities of several fast-fading MIMO
channels with different number of source, destination, and
eavesdropper antennas are plotted against the source signal-to-noise
ratio (SNR) $P/\sigma^2$ where $\sigma_D^2=\sigma_W^2=\sigma^2$. The
channel gain advantage of the eavesdropper is set to $\alpha^2=1$. We
observe that the key capacity levels off as $P/\sigma^2$ increases in
three of the four channels, except the case of
$(m_S,m_D,m_W)=(2,1,1)$, considered in Fig.~\ref{fig1}. It appears
that the relative antenna dimensions determine the asymptotic behavior
of the key capacity when the SNR is large.  To more precisely study
this behavior, we evaluate the limiting value of $C_K$ as the input
power $P$ of the source becomes very large.  To highlight the
dependence of $C_K$ on $P$, we use the notation $C_K(P)$.
\begin{corollary}\label{thm:keycaplimP}
\begin{enumerate}
\item If $m_W \geq m_S$, then 
\[
\lim_{P \rightarrow \infty} C_K(P) = E\left[ \log \frac{\det \left(
      H_W^{\dagger} H_W + \frac{\sigma_W^2}{\alpha^2 \sigma_D^2}
      H_D^{\dagger} H_D \right)}{\det \left(H_W^{\dagger} H_W \right)}
\right].
\]
\item Suppose that $m_W <  m_S$. Define
\[
C_{\infty}(P) = E\left[ \log\det \left( I_{m_D} +
    \frac{P}{m_S\sigma_D^2} H_D \left[I_{m_S} - H_W^{\dagger}
      \left(H_W H_W^{\dagger} \right)^{-1} H_W \right] H_D^{\dagger}
  \right) \right].
\]
Then $\lim_{P \rightarrow \infty} \frac{C_K(P)}{C_{\infty}(P)} = 1$.
\end{enumerate}
\end{corollary}
\begin{proof}
 First fix $(\lambda_1, \lambda_2, \ldots, \lambda_{m_S})
  = \left(\frac{P}{m_S},\frac{P}{m_S}, \ldots, \frac{P}{m_S}\right)$
  or equivalently $K_X=\frac{P}{m_s} I_{m_S}$, and consider the
  mapping $G$ defined in the proof of Lemma~\ref{thm:Elogdet} as a
  function of $P$.  Also define
\[
\hat f(P) = \log\det \left( I_{m_D} + \frac{P}{m_S\sigma_D^2} H_D
  \left(I_{m_S} + \frac{\alpha^2P}{m_S\sigma_W^2} H_W^{\dagger} H_W
  \right)^{-1} H_D^{\dagger} \right).
\]
Thus $C_K(P) = E[\hat f(P)]$.  It is not hard to check that for any $P
< \tilde P$, $G(\tilde P) \geq G(P)$, which implies that $\det(G(P))
\geq \det(G(\tilde P))$. Hence $\hat f$ is increasing in $P$.  Since
the elements of $H_W$ are continuously i.i.d.,
$\mathrm{rank}(H_W^{\dagger} H_W) = \mathrm{rank}(H_W H_W^{\dagger}) =
\mathrm{rank}(H_W) = \min(m_S,m_W)$ w.p.1. Thus the matrix
$H_W^{\dagger} H_W$ (resp. $H_W H_W^{\dagger}$) is invertible w.p.1
when $m_W \geq m_S$ (resp. $m_W < m_S$).

Now, consider the case of $m_W \geq m_S$. As in (\ref{e:CKub}), we have
\[
\hat f(P) = \log \frac{\det \left( \frac{m_S
      \sigma_W^2}{\alpha^2 P} I_{m_S} + H_W^{\dagger} H_W +
    \frac{\sigma_W^2}{\alpha^2 \sigma_D^2} H_D^{\dagger} H_D
  \right)}{\det \left(\frac{m_S \sigma_W^2}{\alpha^2 P}I_{m_S} +
    H_W^{\dagger} H_W \right)}.
\]
Since $H_W^{\dagger} H_W$ is invertible w.p.1, 
\[
\lim_{P \rightarrow \infty} \hat f(P) = \log \frac{\det \left(
    H_W^{\dagger} H_W + \frac{\sigma_W^2}{\alpha^2 \sigma_D^2}
    H_D^{\dagger} H_D \right)}{\det \left(H_W^{\dagger} H_W \right)}
~~~~~\mbox{w.p.1}.
\]
Hence Part 1) of the lemma results from monotone convergence.

For the case of $m_W < m_S$, the matrix inversion formula allows us to
instead write
\[
\hat f(P) = \log\det \left( I_{m_D} + \frac{P}{m_S\sigma_D^2}
  H_D \left[I_{m_S} - H_W^{\dagger} \left( \frac{m_S
        \sigma_W^2}{\alpha^2 P} I_{m_W} + H_W H_W^{\dagger}
    \right)^{-1} H_W \right] H_D^{\dagger} \right)
\]
Since $H_W H_W^{\dagger}$ is invertible w.p.1, 
we can also define
\[
\hat f_{\infty}(P) = \log\det \left(
  I_{m_D} + \frac{P}{m_S\sigma_D^2} H_D \left[I_{m_S} - H_W^{\dagger}
    \left(H_W H_W^{\dagger} \right)^{-1} H_W \right] H_D^{\dagger} \right).
\]
Note that $C_{\infty}(P)=E[\hat f_{\infty}(P)]$.  Since $H_W$ is of
rank $m_W$ w.p.1, it has the singular value decomposition $H_W = U_W
\left[ S_W \ 0_{m_S-m_W} \right] V_W^{\dagger}$, where $S_W =
\mbox{diag}(s_1,s_2,\ldots,s_{m_W})$ is a diagonal matrix whose
diagonal elements are the positive singular values of $H_W$. Also let
$V=[\tilde V \ \hat V]$, i.e., $\tilde V_W$ and $\hat V_W$ consist
respectively of the first $m_W$ and the last $m_S-m_W$ columns of $V$.
Employing the unitary property of $U_W$ and $V_W$, it is not hard to
verify that
\begin{eqnarray}
\hat f(P) &=&
 \log\det \left( I_{m_D}
   + \frac{P}{m_S \sigma_D^2} H_D \hat V_W \hat V_W^{\dagger} H_D^{\dagger} 
   + H_D \tilde V_W \Lambda_W(P) \tilde V_W^{\dagger} H_D^{\dagger} 
 \right)
 \label{e:hfP} \\
\hat f_{\infty}(P) &=&
 \log\det \left( I_{m_D}
   + \frac{P}{m_S \sigma_D^2} H_D \hat V_W \hat V_W^{\dagger} H_D^{\dagger} 
 \right)
 \label{e:hf8P} 
\end{eqnarray}
where $\Lambda_W(P)=\frac{\sigma_W^2}{\alpha^2 \sigma_D^2} \left(
  \frac{m_S \sigma_W^2}{\alpha^2 P} I_{m_W} + S_W^2
\right)^{-1}$. From (\ref{e:hfP}) and (\ref{e:hf8P}), it is clear that
$\hat f_{\infty}(P) \leq \hat f(P)$.

Further let $t(P)=\mathrm{tr} \left(H_D \tilde V_W \Lambda_W(P) \tilde
  V_W^{\dagger} H_D^{\dagger} \right)$. Since $t(P)I_{m_D} \geq H_D
\tilde V_W \Lambda_W(P) \tilde V_W^{\dagger} H_D^{\dagger}$,
\begin{eqnarray}
\hat f(P) &\leq& 
\log\det \left( [1+t(P)]I_{m_D} + \frac{P}{m_S
    \sigma_D^2} H_D \hat V_W \hat V_W^{\dagger} H_D^{\dagger} \right)
\nonumber \\
&=&
m_D \log(1+t(P)) + \log\det \left( I_{m_D} + \frac{P}{m_S
    \sigma_D^2 [1+t(P)]} H_D \hat V_W \hat V_W^{\dagger} H_D^{\dagger} \right).
\label{e:hfPub}
\end{eqnarray}
Let $\mu_1,\mu_2, \ldots, \mu_j$ be the positive eigenvalues of $H_D
\hat V_W \hat V_W^{\dagger} H_D^{\dagger}$. Note that $1\leq j\leq
\min(m_D,m_S-m_W)$, because of the fact that the elements of $H_D$ are
continuously i.i.d. and are independent of the elements of
$H_W$. Hence, from (\ref{e:hf8P}), (\ref{e:hfPub}) and the fact that
$\hat f_{\infty}(P) \leq \hat f(P)$, we have
\begin{eqnarray}
  0 \ \leq \ \hat f(P) - \hat f_{\infty}(P)
  &\leq& 
  m_D \log(1+t(P))+
  \log \left( 
    \frac{\prod_{i=1}^j \left[ 1 + \frac{P\mu_i}{m_S\sigma_D^2 (1+t(P))}
      \right]}{\prod_{i=1}^j \left[ 1 + \frac{P\mu_i}{m_S\sigma_D^2} \right]}
  \right) \nonumber \\
  &=&
  m_D \log(1+t(P)) + \sum_{i=1}^j  \log \left(
  \frac{\frac{1}{1+t(P)} + \frac{m_S\sigma_D^2}{P\mu_i}}{1 + \frac{m_S\sigma_D^2}{P\mu_i}} \right).
\label{e:hf8P-hfP}
\end{eqnarray}

Now note that 
\[
\lim_{P\rightarrow \infty} t(P) = \frac{\sigma_W^2}{\alpha^2
  \sigma_D^2} \mathrm{tr}\left(H_D \tilde V_W S_W^{-2} \tilde
  V_W^{\dagger} H_D^{\dagger} \right) = \frac{\sigma_W^2}{\alpha^2
  \sigma_D^2} \mathrm{tr}\left([H_W^{-1} H_D^{\dagger}]^{\dagger}
  H_W^{-1} H_D^{\dagger} \right)
\]
where $H_W^{-1}$ denotes the Penrose-Moore pseudo-inverse of $H_W$.
Then (\ref{e:hf8P-hfP}) implies that
\begin{eqnarray*}
0 
&\leq&
\liminf_{P\rightarrow \infty} [ \hat f(P) - \hat f_{\infty}(P)]  \\
&\leq&
\limsup_{P\rightarrow \infty} [ \hat f(P) - \hat f_{\infty}(P)] \\
&\leq& 
(m_D-j) \log \left( 1 + \frac{\sigma_W^2}{\alpha^2 \sigma_D^2}
  \mathrm{tr}\left([H_W^{-1} H_D^{\dagger}]^{\dagger} H_W^{-1}
    H_D^{\dagger} \right) \right) \mbox{~~~~~w.p.1.}
\end{eqnarray*}
Hence by Fatou's lemma, we get
\begin{eqnarray}
0 
&\leq&  
\liminf_{P\rightarrow \infty} [ C_K(P) - C_{\infty}(P)] \nonumber \\
&\leq&  
\limsup_{P\rightarrow \infty} [ C_K(P) - C_{\infty}(P)] \nonumber \\
&\leq& 
E\left[
  (m_D-j) \log \left( 1 + \frac{\sigma_W^2}{\alpha^2 \sigma_D^2}
  \mathrm{tr}\left([H_W^{-1} H_D^{\dagger}]^{\dagger} H_W^{-1}
    H_D^{\dagger} \right) \right) \right].
\label{e:CK-C8}
\end{eqnarray}
From (\ref{e:hf8P}), it is clear that $\hat f_{\infty}(P)$ increases
without bound in $P$ w.p.1; hence $C_{\infty}(P)$ also increases
without bound.  Combining this fact with (\ref{e:CK-C8}), we arrive at
the conclusion of Part 2) of the lemma.
\end{proof}
Part 1) of the lemma verifies the observations shown in
Fig.~\ref{fig1} that the key capacity levels off as the SNR increases
if the number of source antennas is no larger than that of
eavesdropper antennas. When the source has more antennas, Part 2) of
the lemma suggests that the key capacity can grow without bound as $P$
increases similarly to a MIMO fading channel with capacity
$C_{\infty}(P)$.  Note that the matrix $I_{m_S} - H_W^{\dagger}
\left(H_W H_W^{\dagger} \right)^{-1} H_W$ in the expression that
defines $C_{\infty}(P)$ is a projection matrix to the orthogonal
complement of the column space of $H_W$. Thus $C_{\infty}(P)$ has the
physical interpretation that the secret information is passed across
the dimensions not observable by the eavesdropper. The most
interesting aspect is that this mode of operation can be achieved even
if neither the source nor the destination knows the channel matrix
$H_W$.

We note that the asymptotic behavior of the key capacity in the high
SNR regime summarized in Corollary~\ref{thm:keycaplimP} is similar to
the idea of secrecy degree of freedom introduced in
\cite{KhistiISIT07}. The subtle difference here is that no up-to-date
CSI of the destination channel is needed at the source.

Another interesting observation from Fig.~\ref{fig1} is that for the
case of $(m_S,m_D,m_W)=(1,10,10)$, the source power $P$ seems to have
little effect on the key capacity. A small amount of source power is
enough to get close to the leveling key capacity of about $1$ bit per
channel use. This observation is generalized below by
Corollary~\ref{thm:keycapant}, which characterizes the effect of
spatial dimensionality of the destination and eavesdropper on the key
capacity when the destination and eavesdropper both have a large
number of antennas.
\begin{corollary} \label{thm:keycapant} 
  When $m_D$ and $m_W$ approaches infinity in such a way that
  $\displaystyle \lim_{m_D, m_W \rightarrow \infty} \frac{m_W}{m_D} = \beta$,
\[
C_K \rightarrow m_S \log \left( 1 + \frac{1}{\beta \alpha^2
    \sigma^2_D/\sigma^2_W} \right).
\]
\end{corollary}
\begin{proof}
  This corollary is a direct consequence of the fact that
  $\frac{1}{m_D}H_D^{\dagger} H_D \rightarrow I_{m_S}$ and
  $\frac{1}{m_W}H_W^{\dagger} H_W \rightarrow I_{m_S}$ w.p.1, which is
  in turn due to the strong law of large numbers.
\end{proof}
Note that we can interpret the ratio $\beta$ as the spatial
dimensionality advantage of the eavesdropper over the destination. The
expression for the limiting $C_K$ in the corollary clearly indicates
that this spatial dimensionality advantage affects the key capacity in
the same way as the channel gain advantage $\alpha^2$.

\begin{figure}
  \begin{center}
   \includegraphics[width=0.75\textwidth]{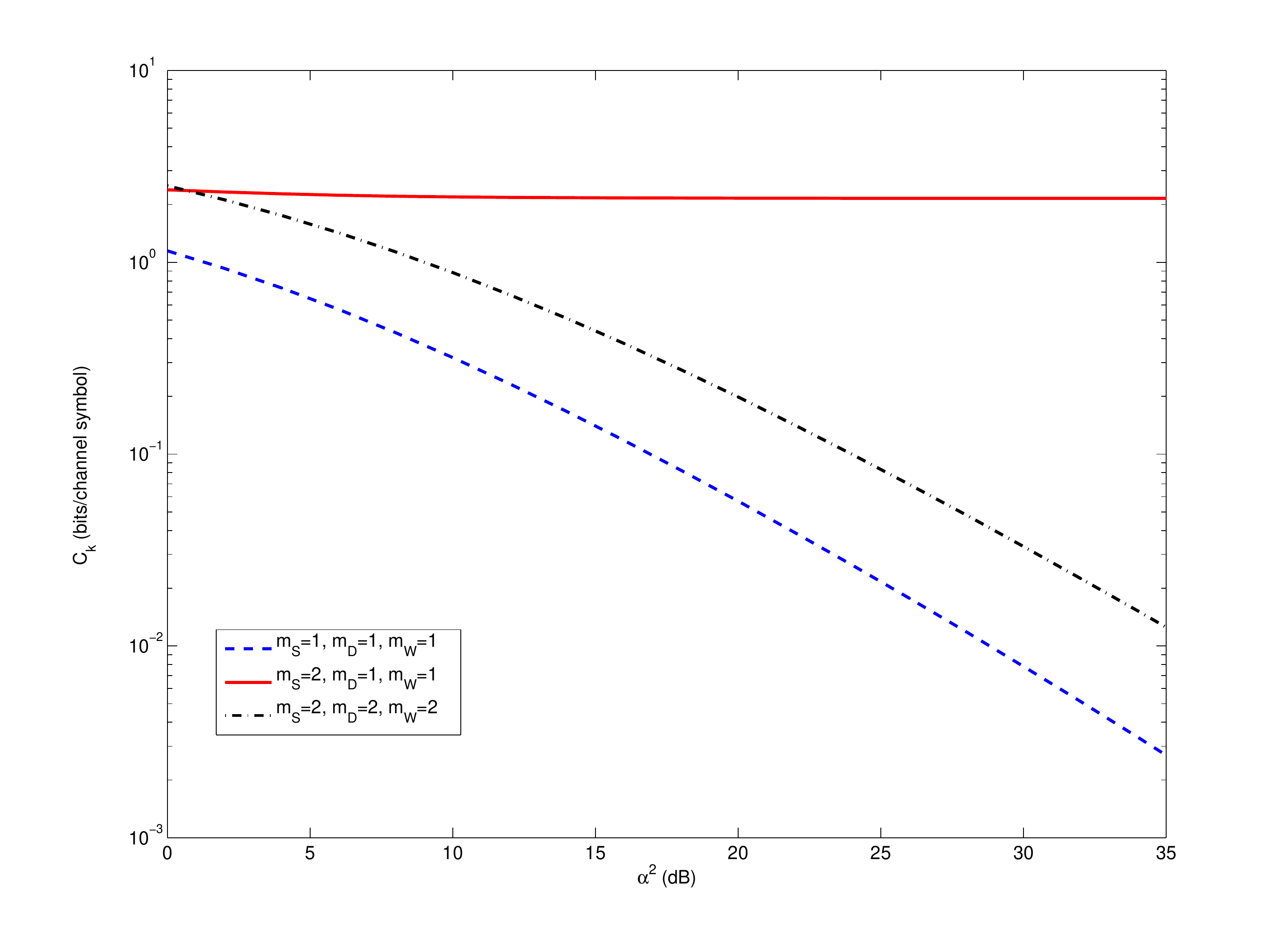}
   \caption{Key capacities of fast-fading MIMO wiretap channels with different numbers of source, destination,
     eavesdropper antennas. The source signal to noise ratio $P/\sigma^2=10$dB, where $\sigma_D^2=\sigma_W^2=\sigma^2$.}
   \vspace{-20pt}
   \label{fig2}
  \end{center}
\end{figure}
In Fig.~\ref{fig2}, the key capacities of several fast-fading MIMO
channels with different numbers of source, destination, and
eavesdropper antennas are plotted against the eavesdropper's channel
gain advantage $\alpha^2$, with $P/\sigma^2=10$dB.  The results in
Fig.~\ref{fig2} show the other effect of spatial dimensionality. We
observe that the key capacity decreases almost reciprocally with
$\alpha^2$ in the channels with $(m_S,m_D,m_W)=(1,1,1)$ and
$(m_S,m_D,m_W)=(2,2,2)$, but stays almost constant for the channel
with $(m_S,m_D,m_W)=(2,1,1)$. It seems that the relative numbers of
source and eavesdropper antennas again play the main role in
differentiating these two different behaviors of the key capacity.  To
verify that, we evaluate the limiting value of $C_K$ as the gain
advantage $\alpha^2$ of the eavesdropper becomes very large. To
highlight the dependence of $C_K$ on $\alpha^2$, we use the notation
$C_K(\alpha^2)$.
\begin{corollary}\label{thm:keycaplim}
$\lim_{\alpha \rightarrow \infty} C_K(\alpha^2) 
=  \left\{
\begin{array}{ll}
  0 & \ \mathrm{if}\ m_W \geq m_S \\
  C_{\infty}(P) & \ \mathrm{if}\ m_W < m_S.
\end{array} \right.$
\end{corollary}
\begin{proof}
  Similar to the proof of Corollary~\ref{thm:keycaplimP}.
\end{proof}
Similar to the case of large SNR, when the number of source antennas
is larger than that of the eavesdropper's antennas, secret information
can be passed across the dimensions not observable by the
eavesdropper. This can be achieved with neither the source nor the
destination knowing the channel matrix $H_W$.

\section{Alternative Achievability of Key Capacity} \label{se:proof}

In this section, we provide an alternative proof of achievability for
key capacity, which does not require the transmission of continuous
symbols over the public channel. We derive the result from ``first
principles'', which provides more insight on the desirable structure
of a practical key agreement scheme. The main steps of the key
agreement procedure are the following:
\begin{enumerate}
\item the source sends a sequence of i.i.d. symbols $X^n$;
\item the destination ``quantizes'' its received sequence $Y^n$ into
  $\hat Y^n$ with a Wyner-Ziv compression scheme;
\item the destination uses a binning scheme with the quantized symbol
  sequences to determine the secret key and the information to feed
  back to the source over the public channel;
\item the source exploits the information sent by the destination to reconstruct the destination's quantized sequence
  $\hat Y^n$ and uses the same binning scheme to generate its secret key.
\end{enumerate}
The secrecy of the resulting key is established by carefully
structuring the binning scheme.

For the memoryless wiretap channel $(X,Y,Z)$ specified by the joint pdf $p(y|x)p(z|x)p(x)$, consider the quadruple
$(X,Y,\hat Y, Z)$ defined by the joint pdf $p(x,y,\hat y,z) = p(\hat y|y)p(y|x)p(z|x)p(x)$ with $p(\hat y|y)$ to be
specified later. We assume that $\hat Y$ takes values in the alphabet $\mathcal{Y}$. Given a sequence of $n$ elements
$x_n=(x_1, x_2, \ldots, x_n)$, $p(x^n)=\prod_{j=1}^{n} p(x_j)$ unless otherwise specified. Similar notation and
convention apply to all other sequences as well as their corresponding pdfs and conditional pdfs considered hereafter.

\subsection{Random Code Generation}
Choose $p(\hat y|y)$ such that $I(X;\hat Y) - I(\hat Y; Z) >0$ and
$I(\hat Y;Z)>0$, and let $p(\hat y)$ denote the corresponding
marginal. Note that the existence of such $p(\hat y|y)$ can be assumed
without loss of generality if $I(X;Y)-I(Y;Z)>0$ and $I(Y;Z)>0$. If
$I(X;Y)-I(Y;Z)=0$, there is nothing to prove. Similarly, if
$I(Y;Z)=0$, the construction below can be trivially modified to show
that $I(X;Y)$ is an achievable key rate.

Fix a small (small enough so that the various rate definitions and
bounds on probabilities below make sense and are non-trivial)
$\varepsilon > 0$.  Let us define
\begin{eqnarray}
  R_1 &\stackrel{\Delta}{=}& I(Y;\hat Y) + 4 \varepsilon \nonumber \\
  R_2 &\stackrel{\Delta}{=}& I(Y;\hat Y) - I(X;\hat Y) + 22\varepsilon \nonumber \\
  R_3 &\stackrel{\Delta}{=}& I(X;\hat Y) - I(\hat Y; Z) - \varepsilon \nonumber \\
  R_4 &\stackrel{\Delta}{=}& I(\hat Y;Z) - 17 \varepsilon.\label{eq:def_rates}
\end{eqnarray}

For each $j=1,2,\ldots, 2^{nR_2}$ and $l=1,2, \ldots, 2^{nR_3}$, generate $2^{nR_4}$ codewords $\hat Y^n(j,l,1), \hat
Y^n(j,l,2), \ldots, \hat Y^n(j,l,2^{nR_4})$ according to $p(\hat y^n)$. The set of codewords $\{\hat Y^n(j,l,k)\}$ with
${k=1\dots2^{nR_4}}$ forms a subcode denoted by $\mathcal{C}(j,l)$. The union of all subcodes $\mathcal{C}(j,l)$ for
$j=1,2,\ldots, 2^{nR_2}$ and $l=1,2, \ldots, 2^{nR_3}$ forms the code $\mathcal{C}$. For convenience, we denote the
$2^{nR_1}$ codewords in $\mathcal{C}$ as $\hat Y^n(1), \hat Y^n(2), \ldots, \hat Y^n(2^{nR_1})$, where $\hat
Y^n(j+(l-1)2^{nR_2}+(w-1)2^{n(R_2+R_3)}) = \hat Y^n(j,l,w)$ for $j=1,2,\ldots, 2^{nR_2}$, $l=1,2, \ldots, 2^{nR_3}$, and
$w=1,2,\ldots, 2^{nR_4}$. The code $\mathcal{C}$ and its subcodes $\mathcal{C}(j,l)$ is revealed to the source,
destination, and eavesdropper. In the following, we refer to a codeword or its index in $\mathcal{C}$
interchangeably. Under this convention, the subcode $\mathcal{C}(j,l)$ is also the set that contains all the indices of
its codewords. Denote $\mathcal{\hat C}(j) = \bigcup_{l=1}^{2^{nR_3}} \mathcal{C}(j,l)$ and $\mathcal{\tilde C}(l) =
\bigcup_{j=1}^{2^{nR_2}} \mathcal{C}(j,l)$.

\subsection{Secret Sharing Procedure}
For convenience, we define the joint typicality indicator function
$T_{\varepsilon}(\cdot)$ that takes in a number of sequences as its
arguments. The value of $T_{\varepsilon}(\cdot)$ is $1$ if the
sequences are $\varepsilon$-jointly typical, and the value is $0$
otherwise. Further define the indicator function for the sequence pair
$(y^n,\hat y^n)$:
\[
S_{\varepsilon}(y^n,\hat y^n) = \left\{
  \begin{array}{ll}
    1 & \ \mathrm{if}\ 
    \Pr\{T_{\varepsilon}(X^n,y^n,\hat y^n,Z^n)=1\} \geq 1-\varepsilon \\
    0 & \ \mathrm{otherwise}
  \end{array} \right.
\]
where $(X^n,Z^n)$ is distributed according to $p(x^n,z^n|y^n,\hat
y^n)$ in the definition above.

The source generates a random sequence $X^n$ distributed according to
$p(x^n)$. If $X^n$ satisfies the average power constraint
(\ref{e:powerconstr}), the source sends $X^n$ through the $(X,Y,Z)$
channel. Otherwise, it ends the secret-sharing process.  Since $p(x)$
satisfies $E[|X|^2]\leq P$, the law of large numbers implies that the
probability of the latter event can be made arbitrarily small by
increasing $n$. Hence we can assume below, with no loss of generality,
that $X^n$ satisfies (\ref{e:powerconstr}) and is sent by the
source. This assumption helps to make the probability calculations in
Section~\ref{se:analysis} less tedious.

Upon reception of the sequence $Y^n$, the destination tries to
quantize the received sequence. Let $M$ be the output of its
quantizer. Specifically, if there is a unique sequence $\hat
Y^n(m)\in\mathcal{C}$ for some $m \in \{1,2,\ldots,2^{nR_1}\}$ such
that $S_{\varepsilon}(Y^n,\hat Y^n(m))=1$, then it sets the output of
the quantizer to $M=m$. If there is more than one such sequence, $M$
is set to be the smallest sequence index $m$. If there is no such
sequence, it sets $M=0$. Let $L$ and $J$ be the unique indices such
that $\hat Y^n(M)\in\mathcal{C}(J,L)$. The index $L$ will be used as
the key while the index $J$ is fed back to the source over the public
channel, i.e. $\Psi_k = J$. If $M=0$, set $J=0$ and choose $L$
randomly over $\{1, 2, \ldots, 2^{nR_3}\}$ with uniform probabilities.

After receiving the feedback information $J$ via the public channel,
the source attempts to find a unique $\hat Y^n(m) \in \mathcal{C}$
such that $T_\varepsilon(X^n,\hat Y^n(m))=1$ and $m \in \mathcal{\hat
  C}(J)$. If there is such a unique $\hat Y^n(m)$, the source decodes
$\hat M = m$. If there is no such sequence or more than one such
sequence, the source sets $\hat M=0$. If $J=0$, it sets $\hat M = 0$.
Finally, if $\hat M >0$, the source generates its key $K=k$, such that
$\hat M \in \mathcal{C}(J,k)$. If $\hat M =0$, it sets $K=0$.

We also consider a fictitious receiver who observes the sequence $Z^n$
and obtains both indices $J$ and $L$ via the public channel. This
receiver sets $\tilde M=0$ if $J=0$. Otherwise, it attempts to find a
unique $\hat Y^n(m) \in \mathcal{C}$ such that $T_\varepsilon(\hat
Y^n(m),Z^n)=1$ and $m \in \mathcal{C}(J,L)$. If there is such a unique
$\hat Y^n(m)$, the source decodes $\tilde{M} = m$. If there is no such
sequence or more than one such sequence, the source sets
$\tilde{M}=0$.

\subsection{Analysis of Probability of Error}
\label{se:analysis}
We use a random coding argument to establish the existence of a code
with rates given by (\ref{eq:def_rates}) such that $\Pr\{K\neq L\}$
and $\Pr\{M\neq\tilde{M}\}$ vanish in the limit of large block length
$n$. Without further clarification, we note that the probabilities of
the events below, except otherwise stated, are over the joint
distribution of the codebook $\mathcal{C}$, codewords, and all other
random quantities involved.

Before we proceed, we introduce the following lemma regarding the
indicator function $S_\varepsilon$.
\begin{lemma} \label{thm:Se}
\begin{enumerate}
\item If $(Y^n,\hat Y^n)$ distributes according to $p(y^n,\hat y^n)$,
  then $\Pr\{S_{\varepsilon}(Y^n,\hat Y^n)=1\} > 1 - \varepsilon$ for
  sufficiently large $n$.
\item If $\hat Y^n$ distributes according to $p(\hat y^n)$, then
  $\Pr\{S_{\varepsilon}(y^n,\hat Y^n)=1\} \leq
  \frac{2^{-n(R_1-7\varepsilon)}}{1-\varepsilon}$ for all $y^n$.
\item If $Y^n$ distributes according to $p(y^n)$, then
  $\Pr\{S_{\varepsilon}(Y^n,\hat y^n)=1\} \leq
  \frac{2^{-n(R_1-7\varepsilon)}}{1-\varepsilon}$ for all $\hat y^n$.
\item If $(Y^n,\hat Y^n)$ distributes according to $p(y^n) p(\hat
  y^n)$, then $\Pr\{S_{\varepsilon}(Y^n,\hat Y^n)=1\} >
  (1-\varepsilon) \cdot 2^{-n(R_1-\varepsilon)}$ for sufficiently
  large $n$.
\end{enumerate}
\end{lemma}
\begin{proof}
\begin{enumerate}
\item This claim is actually shown in~\cite{Oohama1997}. We briefly
  sketch the proof here using our notation for completeness and easy
  reference.  By the reverse Markov inequality~\cite{Oohama1997},
\[  
\Pr\{S_{\varepsilon}(Y^n,\hat Y^n)=1\} \geq 1 -
\frac{1-\Pr\{T_{\varepsilon}(X^n,Y^n,\hat
  Y^n,Z^n)=1\}}{1-(1-\varepsilon)} > 1- \varepsilon
\]
where the second inequality is due to that fact that
$\Pr\{T_{\varepsilon}(X^n,Y^n,\hat Y^n,Z^n)=1\} > 1-\varepsilon^2$ for
sufficiently large $n$.

\item First, we only need to consider typical $y^n$ since the bound is
  trivial when $y^n$ is not typical. Notice that for any such $y^n$,
\begin{eqnarray*}
  1 &\geq &
  \int T_{\varepsilon}(x^n,y^n,\hat y^n,z^n) p(x^n,\hat y^n,z^n|y^n)
  dx^n dz^n d\hat y^n \\
  &=& 
  \int \Pr\{T_{\varepsilon}(X^n,y^n,\hat y^n,Z^n)=1\} \cdot
  \frac{p(y^n,\hat y^n)}{p(y^n)} d\hat y^n \\
  &\geq &
  \int \Pr\{T_{\varepsilon}(X^n,y^n,\hat y^n,Z^n)=1\} \cdot
  \frac{2^{-n(h(Y,\hat Y)+\varepsilon)}}{2^{-n(h(Y)-\varepsilon)}} d\hat y^n \\
  &=& 2^{-n(h(\hat Y|Y)+2\varepsilon)} \int
  \Pr\{T_{\varepsilon}(X^n,y^n,\hat y^n,Z^n)=1\}
  d\hat y^n.
\end{eqnarray*}
Hence
\begin{eqnarray}
2^{n(h(\hat Y|Y)+2\varepsilon)} &\geq&
 \int \Pr\{T_{\varepsilon}(X^n,y^n,\hat y^n,Z^n)=1\} d\hat y^n \nonumber\\
&\geq&
\int S_{\varepsilon}(y^n,\hat y^n) \cdot 
 \Pr\{T_{\varepsilon}(X^n,y^n,\hat y^n,Z^n)=1\} d\hat y^n \nonumber\\
&\geq&
(1-\varepsilon) \int S_{\varepsilon}(y^n,\hat y^n) d\hat y^n.
\label{e:intSe}
\end{eqnarray}
Now
\begin{eqnarray*}
\Pr\{S_{\varepsilon}(y^n,\hat Y^n)=1\} &=&
\int S_{\varepsilon}(y^n,\hat y^n) p(\hat y^n) d\hat y^n \\
&\leq &
\int S_{\varepsilon}(y^n,\hat y^n) 2^{-n(h(\hat Y)-\varepsilon)} d\hat y^n \\
&\leq &
\frac{2^{-n(I(Y;\hat Y)-3\varepsilon)}}{1-\varepsilon},
\end{eqnarray*}
where the last inequality is due to (\ref{e:intSe}).

\item Same as Part 2), interchanging the roles of $y^n$ and $\hat
  y^n$.

\item From Part 1), we get
\begin{eqnarray*}
1-\varepsilon &<& 
\int S_{\varepsilon}(y^n,\hat y^n) p(y^n,\hat y^n) dy^nd\hat y^n \\
&=&
\int S_{\varepsilon}(y^n,\hat y^n) \frac{p(y^n,\hat y^n)}{p(y^n)p(\hat y^n)}
p(y^n)p(\hat y^n)dy^nd\hat y^n \\
&\leq&
\int S_{\varepsilon}(y^n,\hat y^n) \cdot 
\frac{2^{-n(h(Y,\hat Y)-\varepsilon)}}{2^{-n(h(Y)+\varepsilon)}\cdot 
  2^{-n(h(\hat Y)+\varepsilon)}}  \cdot p(y^n)p(\hat y^n)dy^nd\hat y^n \\
&=& 
2^{n(I(Y;\hat Y)-3\varepsilon)} \Pr\{S_{\varepsilon}(Y^n,\hat Y^n)=1\}.
\end{eqnarray*}
\end{enumerate}
\end{proof}

Moreover we need to bound the probabilities of the following events
pertaining to $M$.
\begin{lemma}\label{thm:PMb}
\begin{enumerate}
\item $\Pr\{M=0\} < 2 \varepsilon$ for sufficiently large $n$.
\item For $m=1,2,\ldots,2^{nR_1}$, $\Pr\{M=m\} \leq
  \frac{2^{-n(R_1-7\varepsilon)}}{1-\varepsilon}$.
\item When $n$ is sufficiently large, $\Pr\{M=m\} \geq \left[1 -
    \frac{2^{-n(R_1-7\varepsilon)}}{1-\varepsilon}\right]^{m-1} \cdot
  (1-\varepsilon)2^{-n(R_1-\varepsilon)}$ uniformly for all
  $m=1,2,\ldots,2^{nR_1}$.
\item When $n$ is sufficiently large, $\Pr\{J=j,L=l\} >
  (1-\varepsilon)^4 \cdot 2^{-n(R_1-R_4+6\varepsilon)}$ uniformly for
  all $j=1,2,\ldots,2^{nR_2}$ and $l=1,2,\ldots,2^{nR_3}$.
\end{enumerate}
\end{lemma}
\begin{proof}
\begin{enumerate}
\item We will use an argument similar to the one in the achievability
  proof of rate distortion function in \cite[Section 10.5]{Cover2006}
  to bound $\Pr \{M=0\}$.  First note that $\{M=0\}$ is the event that
  $S_{\varepsilon}(Y^n,\hat Y^n(m))=0$ for all $m \in
  \{1,2,\ldots,R_1\}$, and hence
\begin{eqnarray}
\Pr \{M=0 \} &=&
 \Pr\left\{ \bigcap_{m=1}^{2^{nR_1}} 
 \{ S_{\varepsilon}(Y^n, \hat Y^n(m)) =0\} \right\} \nonumber \\
 &=&
  \int \left[ \Pr\{S_{\varepsilon}(y^n, \hat Y^n(1)) =0\} 
 \right]^{2^{nR_1}} p(y^n) dy^n ,
\label{e:PM0}
\end{eqnarray}
where the second equality is due to the fact that $\hat Y^n(1),
\ldots, \hat Y^n(2^{nR_1})$ are i.i.d.\ given each fixed~$y^n$.
But
\begin{eqnarray}
  \left[ \Pr\{S_{\varepsilon}(y^n, \hat Y^n(1)) =0\}\right]^{2^{nR_1}}
  &=&
  \left[
    1-\int S_{\varepsilon}(y^n,\hat y^n) p(\hat y^n) d\hat y^n
  \right]^{2^{nR_1}} \nonumber \\
  &=&
  \left[
    1-\int S_{\varepsilon}(y^n,\hat y^n) p(\hat y^n|y^n)
    \frac{p(y^n) p(\hat y^n)}{p(y^n,\hat y^n)} d\hat y^n
  \right]^{2^{nR_1}} \nonumber \\
  & \leq &
  \left[
    1- \int S_{\varepsilon}(y^n,\hat y^n) p(\hat y^n|y^n)
    \frac{2^{-n(h(Y)+\varepsilon)} \cdot 2^{-n(h(\hat Y)+\varepsilon)}}%
    {2^{-n(h(Y,\hat Y)-\varepsilon)}} d\hat y^n
  \right]^{2^{nR_1}} \nonumber \\
  &=&
  \left[
    1- 2^{-n(I(Y;\hat Y)+3\varepsilon)} 
    \int S_{\varepsilon}(y^n,\hat y^n) p(\hat y^n|y^n) d\hat y^n
  \right]^{2^{nR_1}} \nonumber \\
  & \leq &
  1 - \int S_{\varepsilon}(y^n,\hat y^n) p(\hat y^n|y^n) d\hat y^n 
  + \exp\left(-2^{n\varepsilon}\right),
 \label{e:PMb1}
\end{eqnarray}
where the inequality on the third line is due to the fact that
$S_{\varepsilon}(y^n,\hat y^n)=1$ implies $T_{\varepsilon}(y^n,\hat
y^n)=1$, and the last line results from the inequality $(1-xy)^k \leq
1-x+e^{-ky}$ for all $0 \leq x,y \leq 1$ and positive integer $k$
\cite[Lemma 10.5.3]{Cover2006}.  Substituting (\ref{e:PMb1}) back into
(\ref{e:PM0}) and using Lemma~\ref{thm:Se} Part~1), we get
\[
\Pr \{M=0\} \leq  
1 - \Pr\{S_{\varepsilon}(Y^n,\hat Y^n)=1\} + \exp\left(-2^{n\varepsilon}\right) 
 < \varepsilon + \varepsilon = 2\varepsilon
\]
for sufficiently large $n$.

\item Notice that for $m=1,2,\ldots,2^{nR_1}$,
\begin{eqnarray}
\Pr\{M=m\} 
  &=& \Pr\{ S_{\varepsilon}(Y^n,\hat Y^n(m))=1, 
  S_{\varepsilon}(Y^n,\hat Y^n(m-1))=0,  \ldots, 
  S_{\varepsilon}(Y^n,\hat Y^n(1))=0 \} \nonumber \\
  &=& \int \Pr\{ S_{\varepsilon}(y^n,\hat Y^n(1))=1\} 
  \left[\Pr\{ S_{\varepsilon}(y^n,\hat Y^n(1))=0\}\right]^{m-1}
  p(y^n) dy^n
\label{e:PM} 
\end{eqnarray}
where the second equality results from the i.i.d.\ nature of $\hat
Y^n(1), \ldots, \hat Y^n(m)$.  Thus we have
\[
\Pr\{M=m\} 
\leq   \Pr\{ S_{\varepsilon}(Y^n,\hat Y^n(1))=1\} 
\leq  \frac{2^{-n(R_1-7\varepsilon)}}{1-\varepsilon},
\]
where the last inequality is due to Part~2) of Lemma~\ref{thm:Se}
since $Y^n$ and $\hat Y^n(1)$ are independent.

\item From (\ref{e:PM}), we have the lower bound
\begin{eqnarray*}
\Pr\{M=m\} &\geq& 
\left[1 - \frac{2^{-n(R_1-7\varepsilon)}}{1-\varepsilon}\right]^{m-1}
  \Pr\{ S_{\varepsilon}(Y^n,\hat Y^n(1))=1 \} \nonumber \\
&\geq& 
\left[1 - \frac{2^{-n(R_1-7\varepsilon)}}{1-\varepsilon}\right]^{m-1}
\cdot (1-\varepsilon)2^{-n(R_1-\varepsilon)}
\end{eqnarray*}
where the first inequality is due to Part~2) of Lemma~\ref{thm:Se},
and the second inequality is from Part~4) of Lemma~\ref{thm:Se} when
$n$ is sufficiently large. Note that the same sufficiently large $n$
is enough to guarantee the validity of the lower bound above for all
$m=1,2,\ldots,2^{nR_1}$.

\item First note that, for $j=1,2,\ldots,2^{nR_2}$ and
  $l=1,2,\ldots,2^{nR_3}$,
\[
\Pr\{J=j,L=l\} = \sum_{m \in \mathcal{C}(j,l)} \Pr\{M=m\} =
\sum_{w=1}^{2^{nR_4}} \Pr
\left\{M=j+(l-1)2^{nR_2}+(w-1)2^{n(R_2+R_3)}\right\}.
\]
Thus applying Part 3) of the lemma, we get
\begin{eqnarray}
\lefteqn{\Pr\{J=j, L=l\}} \nonumber \\
&\geq&
(1-\varepsilon)2^{-n(R_1-\varepsilon)} \cdot 
\sum_{w=1}^{2^{nR_4}} 
\left[1 - \frac{2^{-n(R_1-7\varepsilon)}}{1-\varepsilon}
\right]^{j-1+(l-1)2^{nR_2}+(w-1)2^{n(R_2+R_3)}}
\nonumber\\
&\geq&
(1-\varepsilon)2^{-n(R_1-\varepsilon)} 
\left[1 - \frac{2^{-n(R_1-7\varepsilon)}}{1-\varepsilon}
\right]^{2^{n(R_2+R_3)}}
\frac{1 - \left[ 1 - 2^{-n(R_1-7\varepsilon)}/(1-\varepsilon)
  \right]^{2^{nR_1}}}{1 - \left[ 1 -
    2^{-n(R_1-7\varepsilon)}/(1-\varepsilon) \right]^{2^{n(R_2+R_3)}}} \nonumber \\
& \geq &
(1-\varepsilon)2^{-n(R_1-\varepsilon)} 
\left[1 - \frac{2^{-n(R_4-7\varepsilon)}}{1-\varepsilon} \right] \cdot
\frac{1 - \left[ 1 - 2^{-n(R_1-7\varepsilon)}/(1-\varepsilon)
  \right]^{2^{nR_1}}}{1 - \left[ 1 -
    2^{-n(R_4-7\varepsilon)}/(1-\varepsilon) \right]}
\nonumber \\
&\geq& 
(1-\varepsilon)^2 \cdot 2^{-n(R_1-R_4+6\varepsilon)} 
\left[1 - \frac{2^{-n(R_4-7\varepsilon)}}{1-\varepsilon} \right]
\left[ 1 - \frac{\exp(-2^{7n\varepsilon})}{1-\varepsilon} \right] \nonumber\\
& > &
(1-\varepsilon)^4 \cdot 2^{-n(R_1-R_4+6\varepsilon)}
\label{e:PMJb}
\end{eqnarray}
uniformly for all $j=1,2,\ldots,2^{nR_2}$ and $l=1,2,\ldots,2^{nR_3}$,
when $n$ is sufficiently large. The lower bound on the fourth line of
(\ref{e:PMJb}) above is obtained from the inequality $(1-x)^k \geq
1-kx$ for any $0 \leq x \leq 1$ and positive integer $k$. The lower
bound on the fifth line is in turn based on the inequality $(1-x)^k
\leq e^{-kx}$ for $0 \leq x \leq 1$ and positive integer $k$.
\end{enumerate}
\end{proof}

We first consider the error event $\{K \neq L\}$. Note that
\begin{eqnarray}
  \Pr\{K \neq L\} &=& \Pr\{M=0\} + \Pr \{M>0, K \neq L\} \nonumber \\
  &=& \Pr\{M=0\} +
  \sum_{m=1}^{2^{nR_1}} \Pr \left\{\mathcal{\tilde E}_m \cup \mathcal{E}_m,  M=m\right\} \nonumber \\
  &\leq & \Pr\{M=0\} +
  \sum_{m=1}^{2^{nR_1}} \Pr \left\{\mathcal{\tilde E}_m, M=m\right\} +
  \sum_{m=1}^{2^{nR_1}} \Pr \left\{\mathcal{E}_m, M=m\right\}
\label{e:PKL}
\end{eqnarray}
where $\mathcal{\tilde E}_m$ is the event $\{T_{\varepsilon}(X^n,\hat
Y^n(m))=0\}$, and $\mathcal{E}_m$ is the event that there is an $m'
\in \mathcal{\hat C}(j)$ such that $m \in \mathcal{\hat C}(j)$,
$m'\neq m$, and $T_{\varepsilon}(X^n,\hat Y^n(m'))=1$. From
(\ref{e:PM}), we have
\begin{eqnarray}
\lefteqn{\Pr \left\{\mathcal{\tilde E}_m, M=m\right\} } \nonumber \\
&=&
\Pr \left\{T_{\varepsilon}(X^n,\hat Y^n(m))=0, S_{\varepsilon}(Y^n,\hat Y^n(m))=1, 
  S_{\varepsilon}(Y^n,\hat Y^n(m-1))=0,  \ldots, S_{\varepsilon}(Y^n,\hat Y^n(1))=0
  \right\}  \nonumber \\
& \leq &
\Pr \Big\{T_{\varepsilon}(X^n,Y^n,\hat Y^n(m),Z^n)=0,
  S_{\varepsilon}(Y^n,\hat Y^n(m))=1, \nonumber \\
& & ~~~~~~~~ 
  S_{\varepsilon}(Y^n,\hat Y^n(m-1))=0,  \ldots, S_{\varepsilon}(Y^n,\hat Y^n(1))=0
  \Big\}  \nonumber \\
&=&
\int \left[ \int \Pr \left\{T_{\varepsilon}(x^n,y^n,\hat Y^n(m),z^n)=0,
    S_{\varepsilon}(y^n,\hat Y^n(m))=1\right\} p(x^n,z^n|y^n) dx^n
  dz^n \right]  \nonumber \\
& & ~~~~~~\cdot 
\prod_{m'=1}^{m-1} \Pr\{S_{\varepsilon}(y^n,\hat Y^n(m'))=0\} p(y^n) dy^n  \nonumber \\
&=&
\int \left(\left\{ \int [1-T_{\varepsilon}(x^n,y^n,\hat y^n,z^n)] 
    p(x^n,z^n|y^n,\hat y^n) dx^n dz^n \right\} 
  \cdot S_{\varepsilon}(y^n,\hat y^n) p(\hat y^n) d\hat y^n \right)  \nonumber \\
& & ~~~~~~\cdot 
\prod_{m'=1}^{m-1} \Pr\{S_{\varepsilon}(y^n,\hat Y^n(m'))=0\} p(y^n) dy^n  \nonumber \\
& \leq & \varepsilon \cdot 
\Pr \left\{S_{\varepsilon}(Y^n,\hat Y^n(m))=1, 
  S_{\varepsilon}(Y^n,\hat Y^n(m-1))=0,  \ldots, S_{\varepsilon}(Y^n,\hat Y^n(1))=0
  \right\}  \nonumber \\
&=& \varepsilon\cdot\Pr\{M=m\},
\label{e:PtE}
\end{eqnarray}
where the equality on the fourth line is due to the i.i.d.\ nature of
$\hat Y^n(1), \ldots, \hat Y^n(2^{nR_1})$, the equality on the fifth
line results from the fact that $p(x^n,z^n|y^n)=p(x^n,z^n|y^n,\hat
y^n)$ (since $(X,Z)\rightarrow Y\rightarrow \hat Y$), and the
inequality on the second last line is from the definition of the
indicator function $S_\varepsilon$.

Similarly assuming $m \in \mathcal{\hat C}(j)$, we have from
(\ref{e:PM})
\begin{eqnarray}
  \Pr \{ \mathcal{E}_m, M=m\}
  &\leq &
  \sum_{\tiny \begin{array}{c} m' \in \mathcal{\hat C}(j) \\ m' \neq m \end{array}}
  \Pr\left\{T_{\varepsilon}(X^n,\hat Y^n(m'))=1, S_{\varepsilon}(Y^n,\hat Y^n(m))=1\right\} \nonumber \\
  & = &
  \sum_{\tiny \begin{array}{c} m' \in \mathcal{\hat C}(j) \\ m' \neq m \end{array}}
  \int \Pr\{T_{\varepsilon}(x^n,\hat Y^n(m'))=1\} \cdot
  \Pr\{S_{\varepsilon}(y^n,\hat Y^n(m))=1\} p(x^n,y^n) dx^n dy^n \nonumber \\
  & \leq &
  2^{n(R_1-R_2)} \cdot 2^{-n(I(X;\hat Y)-3\varepsilon)} \cdot
  \frac{2^{-n(R_1-7\varepsilon)}}{1-\varepsilon} 
  = \frac{2^{-n(R_1+8\varepsilon)}}{1-\varepsilon},
\label{e:PE}
\end{eqnarray}
where the equality on the second line is due to the independence
between $\hat Y^n(m')$ and $\hat Y^n(m)$, and the last inequality
results from Part~2) of Lemma~\ref{thm:Se} and the bound $\Pr\{
T_{\varepsilon}(x^n,\hat Y^n(m')) =1 \} \leq 2^{-n(I(X;\hat
  Y)-3\varepsilon)}$, which is a direct result of \cite[Theorem
15.2.2]{Cover2006}. Hence, substituting the bounds in (\ref{e:PtE})
and (\ref{e:PE}) back into (\ref{e:PKL}) and using Part~1) of
Lemma~\ref{thm:PMb}, we obtain
\begin{equation}
  \Pr\{K \neq L\} \leq 2\varepsilon + \varepsilon \cdot 
  \sum_{m=1}^{2^{nR_1}} \Pr\{M=m\} +
  \sum_{m=1}^{2^{nR_1}} \frac{2^{-n(R_1+8\varepsilon)}}{1-\varepsilon}
  = 2\varepsilon + \varepsilon + \frac{2^{-8n\varepsilon}}{1-\varepsilon}
  < 4\varepsilon
  \label{e:PKLb}
\end{equation}
for $n$ is sufficiently large.

Next we consider the event $\{M\neq\tilde{M}\}$. Define
$\mathcal{\tilde F}_m$ as the event $\{T_{\varepsilon}(\hat
Y^n(m),Z^n)=0\}$ and $\mathcal{F}_m$ as the event that there is an $m'
\in \mathcal{C}(l,j)$ such that $m \in \mathcal{C}(l,j)$, $m' \neq m$,
and $T_{\varepsilon}(\hat Y^n(m'),Z^n)=1$.  Then we have, when $n$ is
sufficiently large, uniformly for all $j=1,2,\ldots,2^{nR_2}$ and
$l=1,2,\ldots,2^{nR_3}$,
\begin{eqnarray}
\lefteqn{\Pr \{\tilde M \neq M | J=j, L=l\}} \nonumber \\
&\leq& 
\sum_{m \in \mathcal{C}(j,l)} \Pr \left\{\mathcal{\tilde F}_m, M=m | 
  J=j, L=l\right\} +
\sum_{m \in \mathcal{C}(j,l)} \Pr \left\{\mathcal{F}_m, M=m |J=j,
    L=l\right\} \nonumber \\
&\leq &
\sum_{m\in \mathcal{C}(j,l)} \varepsilon \cdot \Pr \{M=m | J=j, L=l\} +
\sum_{m\in \mathcal{C}(j,l)}
\frac{2^{-n(R_1+7\varepsilon)}}{1-\varepsilon} \cdot 
\frac{1}{\Pr\{J=j, L=l\}} \nonumber \\
& \leq & 
\varepsilon + \frac{2^{-n(R_1+7\varepsilon)}}{1-\varepsilon} \cdot 
\frac{2^{nR_4}}{(1-\varepsilon)^4 \cdot 2^{-n(R_1-R_4+6\varepsilon)}}
\nonumber \\
& = & 
\varepsilon + \frac{2^{-n\varepsilon}}{(1-\varepsilon)^5} 
< 2\varepsilon.
\label{e:PMJLb}
\end{eqnarray}
Note that the inequality on the third line of (\ref{e:PMJLb}) results
from upper bounds of $\Pr \{\mathcal{\tilde F}_m, M=m\}$ and $\Pr
\{\mathcal{F}_m, M=m\}$, which can be obtained in ways almost
identical to the derivations in (\ref{e:PtE}) and (\ref{e:PE})
respectively. The inequality on the fourth line is, on the other hand,
due to Part 4) of Lemma~\ref{thm:PMb}.

By expurgating the random code ensemble, 
we obtain the following lemma.
\begin{lemma} \label{thm:exist} For any $\epsilon>0$ and $n$
  sufficiently large, there exists a code $\mathcal{C}_n$ with the
  rates $R_1$, $R_2$, $R_3$, and $R_4$ given by (\ref{eq:def_rates})
  such that
\begin{enumerate}
\item $\Pr\{K\neq L | \mathcal{C} = \mathcal{C}_n\} < 8\varepsilon$,
\item $\Pr\{M\neq \tilde{M} | \mathcal{C} = \mathcal{C}_n\} < 8\varepsilon$,
\item $\Pr\{M=m | \mathcal{C} = \mathcal{C}_n\}\leq 
  \frac{2^{-n(R_1-7\varepsilon)}}{1-\varepsilon}$ for all
  $m=1,2,\ldots,2^{nR_1}$, and
\item $\Pr\{L=l | \mathcal{C} = \mathcal{C}_n\} <
  2^{-n(R_3-8\varepsilon)}$ for all $l=1,2,\ldots,2^{nR_3}$.
\end{enumerate}
\end{lemma}

\begin{proof}
Combining Part 1) of Lemma~\ref{thm:PMb}, (\ref{e:PKLb}), and
(\ref{e:PMJLb}), we have
\[
\Pr\{M=0\} + \Pr\{K\neq L\} + \Pr\{M\neq \tilde{M}\} < 8\varepsilon
\]
for sufficiently large $n$. This implies that there must exist a
$\mathcal{C}_n$ satisfying $\Pr\{K\neq L | \mathcal{C} =
\mathcal{C}_n\} < 8\varepsilon$, $\Pr\{M\neq \tilde{M} | \mathcal{C} =
\mathcal{C}_n\} < 8\varepsilon$, and $\Pr\{M=0 | \mathcal{C} =
\mathcal{C}_n\} < 8\varepsilon$. Thus, Parts 1) and 2) are proved.

Now, fix this $\mathcal{C}_n$. For $m=1,2,\ldots,2^{nR_1}$, let $\hat
y^n(m)$ be the $m$th codeword of $\mathcal{C}_n$. Then, by Part~3) of
Lemma~\ref{thm:Se}, 
\[
\Pr\{M=m | \mathcal{C} = \mathcal{C}_n\} \leq \Pr\{
S_{\varepsilon}(Y^n,\hat y^n(m))=1\} \leq
\frac{2^{-n(R_1-7\varepsilon)}}{1-\varepsilon};
\]
hence, Part 3) results.

Note that, for $l=1,2,\ldots,2^{nR_3}$,
\begin{equation}
  \Pr\{L=l|\mathcal{C} = \mathcal{C}_n\} = \Pr\{L=l|M=0,\mathcal{C} =
  \mathcal{C}_n \} \Pr\{M=0|\mathcal{C} = \mathcal{C}_n\} +
  \Pr\{L=l,M>0 | \mathcal{C} = \mathcal{C}_n\}.
\label{e:PL}
\end{equation}
We know from the discussion above that $\Pr\{L=l|M=0,\mathcal{C} =
\mathcal{C}_n \} \Pr\{M=0|\mathcal{C} = \mathcal{C}_n\} < 2^{-nR_3}
\cdot 8\varepsilon$.  Also from Part~3) of
the lemma,
\[
\Pr\{L=l,M>0 |\mathcal{C} = \mathcal{C}_n\} = 
\sum_{m \in \mathcal{\tilde C}_n(l)} \Pr\{M=m | \mathcal{C} =
\mathcal{C}_n\} 
\leq 
 2^{n(R_1-R_3)} \cdot \frac{2^{-n(R_1-7\varepsilon)}}{1-\varepsilon}
= \frac{2^{-n(R_3-7\varepsilon)}}{1-\varepsilon}.
\]
Putting these back into (\ref{e:PL}), we get
\[
  \Pr\{L=l|\mathcal{C} = \mathcal{C}_n\} < 2^{-n(R_3-7\varepsilon)}
  \left[ 8\varepsilon \cdot 2^{-7n\varepsilon}+\frac{1}{1-\varepsilon} \right]
  < 2^{-n(R_3-8\varepsilon)}
\]
for sufficiently large $n$. Thus, Part 4) is proved.
\end{proof}

In the remainder of the paper, we use a fixed code $\mathcal{C}_n$
identified by Lemma~\ref{thm:exist}. For convenience, we drop the
conditioning on $\mathcal{C}_n$.

\subsection{Secrecy Analysis}
\label{sec:secrecy-analysis}
 First we proceed to bound $H(K)$. Note that
 \begin{eqnarray}
 H(K) &=& H(L) + H(K|L) - H(L|K) \nonumber \\
 &\geq& H(L) - H(L|K).
 \label{e:HK}
 \end{eqnarray}
 Using Part~1) of Lemma~\ref{thm:exist} together with Fano's inequality
 gives $H(L|K) \leq 1 + 8n\varepsilon R_3$.  Moreover Part~4) of
 Lemma~\ref{thm:exist} implies that $H(L)> n(R_3-8\varepsilon)$.
 Putting these bounds back into (\ref{e:HK}), we have
 \begin{equation}
   R_3 - (8R_3+8)\varepsilon  - \frac{1}{n} < \frac{1}{n} H(K) \leq R_3.
 \label{e:HKb}
 \end{equation}

Next we bound $I(K;Z^n,J)$. Note that
\begin{eqnarray}
I(K;Z^n,J) &=& I(L;Z^n,J)+I(K;Z^n,J|L) - I(L;Z^n,J|K) \nonumber \\
& \leq &  I(L;Z^n,J)+I(K;Z^n,J|L) \nonumber \\
& \leq &  I(L;Z^n,J)+H(K|L)\nonumber \\
& \leq &  I(L;Z^n,J)+ 8n\varepsilon R_3 + 1
\label{e:IKZJ}
\end{eqnarray}
where the last inequality is obtained from Part~1) of
Lemma~\ref{thm:exist} and Fano's inequality like before. In addition,
it holds that
\begin{eqnarray*}
  I(L;Z^n,J) &=& H(L)-H(L|Z^n,J)\\
  &=& H(L) - H(L,J|Z^n)+H(J|Z^n)\\
  &=& H(L) +H(J|Z^n) - H(L,J,M|Z^n)+H(M|Z^n,L,J)\\
  &\leq& H(L)+H(J) -H(M|Z^n) - {H(L,J|M,Z^n)}+{H(M|Z^n,L,J)}\\
  &\leq& H(L)+H(J) +I(M;Z^n)-H(M) +8nR_1\varepsilon+1,
\end{eqnarray*}
where the second last inequality follows from $H(J|Z^n)\leq H(J)$, and
the last inequality follows from ${H(L,J|M,Z^n)}=0$ (by definition of
$J$ and $L$) and $H(M|Z^n,L,J)\leq 1+8nR_1\varepsilon$ (by Fano's
inequality applied to the fictitious receiver). By construction of the
code $\mathcal{C}_n$, it holds that $H(L)\leq nR_2$ and $H(J) \leq
nR_3$.  In addition, Part 3) of Lemma~\ref{thm:exist} implies
$H(M)\geq n(R_1 -8\varepsilon)$.  Finally, note that $I(M;Z^n)\leq
I(Y^n;Z^n)=nI(Y;Z)$ by the data-processing inequality applied to the
Markov chain $\hat Y^n\rightarrow Y^n\rightarrow Z^n$ and the
memoryless property of the channel between $Y^n$ and $Z^n$. Combining
these observations and substituting the values of $R_1$, $R_2$, and
$R_3$ given by (\ref{eq:def_rates}) back into (\ref{e:IKZJ}), we
obtain
\begin{eqnarray*}
  \frac{1}{n}I(K;Z^n,J) &\leq&
  R_2+R_3-R_1 + I(Y;Z) + (8R_1+8R_3+8)\varepsilon + \frac{2}{n} \\
  &\leq & I(Y;Z)-I(\hat{Y};Z)+ (8R_1+8R_3+9)\varepsilon,
\end{eqnarray*}
when $n$ is sufficiently large.  Without any rate limitation on the
public channel, we can choose the transition probability $p(\hat y|y)$
such that $I(Y;Z)-I(\hat{Y};Z)\leq \varepsilon$; therefore,
\begin{equation}
\frac{1}{n}I(K;Z^n,J)\leq  (8R_1+8R_3+9)\varepsilon.
\label{e:IKZJb}
\end{equation}
Since $\varepsilon>0$ can be chosen arbitrarily, Part~1) of
Lemma~\ref{thm:exist}, (\ref{e:HKb}), and (\ref{e:IKZJb}), establish
the achievability of the secret key rate $I(Y;X)-I(Y;Z)$.

\section{Conclusion}
We evaluated the key capacity of the fast-fading MIMO wiretap
channel. We found that spatial dimensionality provided by the use of
multiple antennas at the source and destination can be employed to
combat a channel-gain advantage of the eavesdropper over the
destination.  In particular if the source has more antennas than the
eavesdropper, then the channel gain advantage of the eavesdropper can
be completely overcome in the sense that the key capacity does not
vanish when the eavesdropper channel gain advantage becomes
asymptotically large. This is the most interesting observation of this
paper, as no eavesdropper CSI is needed at the source or destination
to achieve the non-vanishing key capacity.

\section*{Acknowledgment}
This work was supported by the National Science Foundation under grant number CNS-0626863 and by the Air Force Office of
Scientific Research under grant number FA9550-07-10456. We would also like to thank Dr. Shlomo Shamai and the anonymous
reviewers for their detailed comments and thoughtful suggestions. We are grateful to the reviewer who pointed out a
significant oversight in the proof of Theorem~\ref{thm:keycapgen} in the original version of the paper. We are also
indebted to another reviewer who suggested the concavity argument in the proof of Lemma~\ref{thm:Elogdet}, which is much
more elegant than our original one.




\end{document}